\documentclass[journal]{IEEEtran}
\usepackage{cite}
\usepackage{amsmath,amssymb,amsfonts}
\usepackage{algorithmic}
\usepackage{graphicx}
\usepackage{textcomp}
\usepackage{stfloats}
\usepackage{booktabs}
\usepackage{array}
\usepackage{mathtools}
\usepackage{float}
\usepackage{bm} 
\usepackage{amsmath} 
\usepackage{amssymb}
\usepackage{colortbl}
\usepackage{color} 
\begin{document}

\title{A Comprehensive Study and Comparison of Core Technologies for MPEG 3D Point Cloud Compression}

\author{Hao~Liu,\IEEEmembership{}
        Hui~Yuan,~\IEEEmembership{Senior Member,~IEEE,}
        Qi~Liu,\IEEEmembership{}
        Junhui~Hou,~\IEEEmembership{Member,~IEEE,}
        Ju~Liu,~\IEEEmembership{Senior Member,~IEEE}

\thanks{This work was supported in part by the National Key R{\rm{\&}}D Program of China under Grants 2018YFC0831003; in part by the National Natural Science Foundation of China under Grants 61571274 and 61871342; in part by the open project program of state key laboratory of virtual reality technology and systems, Beihang University, under Grant VRLAB2019B03; in part by the Shandong Natural Science Funds for Distinguished Young Scholar under Grant JQ201614; in part by the Shandong Provincial Key Research and Development Plan under Grant 2017CXGC1504; in part by the Young Scholars Program of Shandong University (YSPSDU) under Grant 2015WLJH39. \emph{Corresponding author: Hui Yuan}.}
\thanks{ H. Liu, Q. Liu, and J. Liu are with  the School of Information Science and Engineering, Shandong University, Qingdao, China, 266200 (e-mail: liuhaoxb@gmail.com; sdqi.liu@gmail.com; juliu@sdu.edu.cn).}
\thanks{H. Yuan (corresponding author) is with the School of Control Science and Engineering, Shandong University, Jinan, China, 250000 (e-mail: huiyuan@sdu.edu.cn).}
\thanks{J. Hou is with the Department of Computer Science, The City University of Hong Kong, Hong Kong, China, 999077 (e-mail: jh.hou@cityu.edu.hk).}
\thanks{This paper includes an additional appendix pdf file which provides rate distortions performance of TMC13 and TMC2 under all experiment settings in details}}

\markboth{IEEE TRANSACTIONS ON BROADCASTING,~Vol.~xx, No.~xx, xx~2019}%
{Shell \MakeLowercase{\textit{et al.}}: Bare Demo of IEEEtran.cls for IEEE Journals}

\maketitle
\begin{abstract}
Point cloud based 3D visual representation is becoming popular due to its ability to exhibit the real world in a more comprehensive and immersive way. However, under a limited network bandwidth, it is very challenging to communicate this kind of media due to its huge data volume. Therefore, the MPEG have launched the standardization for point cloud compression (PCC), and proposed three model categories, i.e., TMC1, TMC2, and TMC3. Because the 3D geometry compression methods of TMC1 and TMC3 are similar, TMC1 and TMC3 are further merged into a new platform namely TMC13. In this paper, we first introduce some basic technologies that are usually used in 3D point cloud compression, then review the encoder architectures of these test models in detail, and finally analyze their rate distortion performance as well as complexity quantitatively for different cases (i.e., lossless geometry and lossless color, lossless geometry and lossy color, lossy geometry and lossy color) by using 16 benchmark 3D point clouds that are recommended by MPEG. Experimental results demonstrate that the coding efficiency of TMC2 is the best on average (especially for lossy geometry and lossy color compression) for dense point clouds while TMC13 achieves the optimal coding performance for sparse and noisy point clouds with lower time complexity.
\end{abstract}


\IEEEpeerreviewmaketitle
\section{Introduction}
\IEEEPARstart{N}{owadays}, 3D models are becoming more and more popular and important in many application fields such as 3D gaming, animation, virtual/augmented reality (VR/AR), and scientific visualization. 3D polygonal meshes can represent a 3D model by reconstructing the 3D surface through polygonal meshes consisting of a set of vertices (geometry) and the corresponding connectivity information (topology). However, there are also lots of natural scenes or objects (such as hairs, forest) with non-manifold geometry [1] which cannot be easily represented by 3D polygonal meshes. Besides, the data volume of a 3D polygonal mesh is very large, while the rendering complexity is also very high. Previously, there are lots of compression methods for 3D polygonal meshes [2]-[8], but they mainly focus on the computer generated (or man-made) 3D models.

In contrast, 3D point clouds are usually composed of a set of 3D coordinates indicating the locations of points, along with some attributes, e.g., color and normal vector. 3D point clouds can represent the points in 3D scenes and objects directly, and thus, the polygonal overhead is saved. Besides, 3D point clouds are flexible enough to represent non-manifold structure, and they are more convenient for computing, transmission, and storage for real time acquired 3D scenes without the connectivity constraint. In the early days, because of the limited 3D sense technology, the 3D acquisition devices can only get the sparse 3D point clouds of small objects, which limited their extensive applications. With the advancement of 3D acquisition technology in recent years, the acquisition of dense 3D point clouds can be accomplished, and thus, the application of 3D point clouds in daily life is becoming more and more applicable, e.g., robotics [9], vehicle-based mobile mapping system [10], intelligent transportation systems [11], VR/AR [12], cultural heritage preservation, automated driving.

Although the 3D point cloud is very practical, the huge data volume of a 3D point cloud limits its extensive applications. Taking a 3D point cloud with one million points of 3D coordinates (12bits per dimension of a position) and RGB (8bits per component of a point) color attributes as an example, the date volume is $1000000{\rm{ \times (12 \times 3 + 8 \times 3)  =  6 \times 1}}{{\rm{0}}^7} \approx 57$M bits. Note that it is only a static point cloud. For a dynamic point cloud with 25 frames per second, the data volume to be transmitted in a second will be $25{\rm{ \times }}57 \approx 1425$M bits, which is unaffordable for the current network bandwidth. Therefore, efficient compression for 3D point cloud is becoming an urgent task to be addressed.

Currently, some LiDAR sensor data compression standards [13][14] have been published and applied in related industries, but they directly compress the geometry raw data (e.g., *.las, *.e57) associated with other attributes such as GPS timestamps, longitude and latitude. Nevertheless, a lot of application fields need fused data that can not be captured by a single sensor. For example, VR/AR needs spatial geometry information from LiDAR and color information captured by RGB cameras to accurately represent colorful 3D scenes. In other words, the geometry and color information should be fused together before application.

Based on the fused data that is usually represented by the Polygon File Format (*.ply), MPEG called for proposals (CfP) from 2017 and is developing the corresponding compression standards for 3D point clouds. In October 2017, the first several platforms, i.e., test model category 1 (TMC1) [15] for static 3D point cloud compression, test model category 2 (TMC2) [16] for dynamic 3D point cloud compression, and test model category 3 (TMC3) [17] for dynamically acquired 3D point clouds, were proposed. Although Schwarz \emph{et al}. [1] have introduced the progress of 3D point cloud compression standardization of MPEG, only qualitative evaluations of the three test models are provided. Since MPEG PCC standards is still an ongoing project, more and more new technologies are emerging. Recently, because the 3D geometric compression methods of TMC1 and TMC3 are similar, TMC1 and TMC3 have been combined into a new platform, namely TMC13 [18] (also called as G-PCC, i.e., geometry-based PCC), while the TMC2 is called as V-PCC (i.e., video-based PCC).

In this paper, we first introduce some basic technologies for 3D point cloud compression briefly, and then introduce the core technologies and the development trends of the G-PCC and V-PCC platforms in detail. After that, we compare the subjective and objective quality of the G-PCC and V-PCC in different experimental settings (i.e., both geometry and color are compressed losslessly; geometry is compressed losslessly, while the color is compressed lossily; both geometry and color are lossily compressed), and summarize some important conclusions based on the experimental results, aiming at encouraging more researchers to participate into the research of 3D point cloud compression.

The remaining of the paper is organized as follows. Section II summarizes the related works of point cloud compression in recent years. Important basic technologies for point cloud compression are briefly presented in Section III. In Section IV, the core technologies of G-PCC and V-PCC are reviewed in detail. Experimental results and analysis are given in Section V, while the conclusions are given in Section VI.

\vspace{3ex}
\section{State Of The Art}

\vspace{3ex}

In the past few years, great progress on 3D point cloud compression has been made [19]-[38]. There are two main problems in point cloud compression. i.e., geometry and attributes (i.e., generally, it refers to color) compression.

For geometry compression, Schnabel and Klein proposed an octree based representation [19] which has been proven to be an efficient way to compress the geometry information. Elseberg \emph{et al}. [20] proposed an efficient octree to store and compress 3D data without loss of precision. Huang \emph{et al}. [21] proposed a progressive point cloud compression based on octree. Besides octree representation, there are also some other lossy compression methods [22]-[24] for geometry information. Ochotta and Saupe [22] partitioned the point cloud in a number of point clusters. A surface patch associated to each cluster is parameterized as a height field, which is efficiently encoded with a shape adaptive wavelet coder. Ahn \emph{et al}. [23] proposed an adaptive range image coding algorithm for the geometry compression of large-scale 3D point clouds. Recently, Rente \emph{et al}. [24] proposed a hybrid geometry compression algorithm based on octree and graph transform.

\vspace{-4.5ex}

Besides geometry information, color information should also be compressed efficiently. Different from geometry information, lossy compression is usually used for color information. Similar to the traditional image compression, the compression of color in 3D point clouds is also composed of prediction, transform, quantization, and entropy coding. With the help of the excellent compression efficiency of JPEG codec, Rufael \emph{et al}. [25] proposed to compress the color information by mapping it to a 2D grid with a depth-first octree traversal. Houshiar and N{\"u}chter [26] mapped the point cloud to the panoramic image, and then compressed the mapped image using the existing image compression methods. This kind of method cannot exploit the inherent correlation among points efficiently. Accordingly, a lot of researchers focused on how to design efficient decorrelation method for the irregular data structure of 3D point clouds. The recently emerged graph transform (GT) [27] [28] has proven to be very suitable for 3D point cloud compression. In the graph transform, a graph is created by calculating the spatial distance of different nodes, and was used to remove color information correlation. Shao \emph{et al}. [29] used a k-d tree partitioning method to code color information based on GT. On the basis of [29], they also proposed a slice-partition schemes and established a tree prediction mode based on k-d tree structure in [30]. To explore the relevance of point cloud attributes, a region adaptive hierarchical transform (RAHT) was proposed in [31]. Besides, sparse representations have also proved to be efficient for 3D point cloud compression. Hou \emph{et al}. [32] proposed a sparse representation method to encode point cloud attributes based on GT basis. Subsequently, they improved the performance of [32] through an inter-block prediction scheme and effective entropy coding strategy in [33]. The above methods are all designed for static point clouds. Since dynamic point clouds are becoming more and more necessary in practical applications, efficient compression methods for dynamic point clouds are also researched. Anis \emph{et al}. [34] used consistent sub-divisional triangular meshes to represent point clouds and design efficient wavelet transforms. Thanou, Chou, and Frossard [35] extended the GT-based work to dynamic point clouds. Queiroz and Chou [36] proposed a motion-compensated approach to encoding dynamic voxelized point clouds. Li \emph{et al}. [37] proposed a novel geometry-based motion prediction method for 3D point clouds. Besides coding tools, efficient rate distortion optimization can also affect the compression efficiency of 3D point clouds. In [38], Liu \emph{et al}. developed a model-based joint color-geometry bit allocation method which uses an octree depth and the JPEG quality value as the encoding parameters.

\begin{figure}[h]
\centering
\includegraphics[width=8.5cm,height=3.7cm]{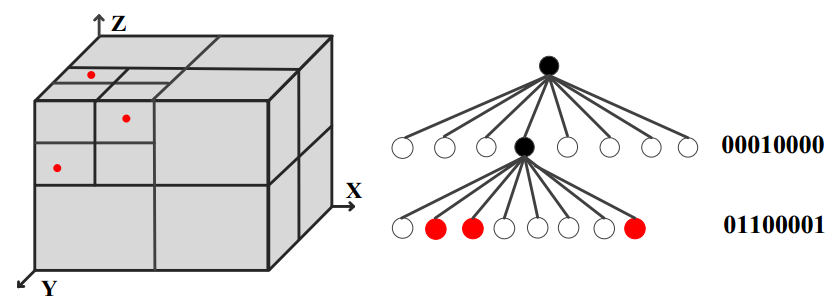}
\caption{Octree decomposition$.$ Current octree depth is 2, three red occupied points are divided into upper left corner in bounding box.}
\end{figure}

\section{Important Basic Technologies}
In 3D point cloud compression, there are some extensively adopted important basic technologies, i.e., octree decomposition, k-d tree decomposition, and level of details (LoD) description. In this section, we will briefly introduce these technologies. Note that octree, k-d tree and LoD are just the preprocess before compression and not  really compress the data. They add overhead for accessing the data. Thus, data size after k-d tree or LoD description could be larger.
\subsection{Octree Decomposition}
In octree-based decomposition, the corresponding position of a cubical bounding box $\mathbb{B}$ is given by two extreme axes (0,0,0) and $(2^n,2^n,2^n)$:
\begin{equation}
\setlength{\abovedisplayskip}{5pt}
\setlength{\belowdisplayskip}{5pt}
{2^n} \ge \max {(\max ({x_i}),\max ({y_i}),\max ({z_i}))_{i = 0,...,T - 1}},
\end{equation} where \textit{n} is the minimum integer satisfying the above inequality, ${({x_i},{y_i},{z_i})_{i = 0,...,T - 1}}$ is the set of 3D positions associated with the points of the original point cloud, and $T$ is the total number of points in point clouds.

As shown in Fig. 1, an octree data structure is generated by splitting $\mathbb{B}$ recursively. At each octree level, a cube (corresponding to a node in the octree) is divided into 8 sub-cubes. A subdivision code with 8 bits is then generated by associating a 1-bit value with each sub-cube in order to indicate whether it contains points (labeled as 1) or not (labeled as 0). Only the sub-cubes that consist of points more than 1 are further divided, until all the sub-cubes only contain 1 point or the predefined octree depth is reached.

\vspace{-2ex}
\subsection{K-d tree Splitting and Nearest Neighbor Search}
The k-d tree is a binary tree in which each leaf node consists of a set of points with \textit{k}-dimensions. k-d tree is widely used in various fields, e.g., nearest neighbor search [39] and collision detection [40]. To take a 3D point cloud as an example, the k-d tree is usually generated based on the three dimensional coordinates of points. At first, the whole point cloud is considered as a root node. Then, a plane is used to divide the root node into two parts. Points belonging to the left (resp. right) of this plane are considered as the left (resp. right) sub-node of the root node. The above splitting method is then iterated based on the left and the right sub-nodes respectively, until a stopping condition is achieved.

In practice, a bounding box containing all the points is generated first.\begin{figure}[h]
\centering
\includegraphics[width=8.5cm,height=4cm]{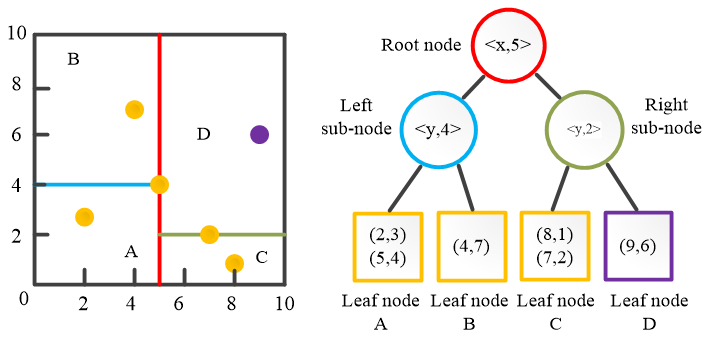}
\caption{2D k-d tree splitting and nearest neighbor search. A 2D bounding box with size $10 \times 10$ is generated, the longest edge of within the bounding box is used to the splitting direction, and the median of the points within the bounding box is applied to be separation point. Assuming that the coordinates of the point being queried is (9,7), the coordinates of the points at which the k-d tree is generated are (2,3),(4,7),(5,4),(9,6),(8,1),(7,2). m indicates splitting direction and n represents the coordinate of separation point in $<$m,n$>$ $.$ The 2D points are divided by k-d tree in yellow and purple rectangle, and (9, 6) is the point closest to the query point in purple rectangles.}
\end{figure} Then, a plane is chosen to divide the bounding box into two sub-bounding box (sub-nodes) along with a certain direction. There are two main methods to choose the splitting direction, one is to divide each dimension of bounding box recursively in a certain order; another is to split the longest edge of the bounding box. After that, the median or the average value of the points' coordinates in the splitting direction is used as the separation position. The splitting strategy is then looped until a termination condition (the number of points in a bounding box is smaller than a predefined threshold or a predefined tree depth, etc.) is achieved. Note that the partitioning method of the k-d tree is likely to be flexible for various application scenarios.

Based on the generated k-d tree, given a target point position, one can easily find its nearest neighbor in a leaf node by a binary tree search operation. Fig. 2 shows a simple example of k-d tree splitting and nearest neighbor search for a set of points with only 2 dimensions.

\subsection{LoD Description}
In the LoD generation, points will be divided into \textit{S} reorganized level sets ${R_s}$ $(s \in \{ 1,...,S\} )$ based on a user specified set of point-to-point spatial Euclidean distance thresholds \textit{dist}$_s$. Note that the distance thresholds must satisfy the following constraints:

\begin{equation} \left\{ \begin{array}{l}
{dist_s} < {dist_{s - 1}},\\
{dist_{S} = 0.}
\end{array}\right.
\end{equation}

As shown in Fig. 3, in the beginning, all the points except for the selected initial point are marked as unvisited, i.e., the set of visited points, denoted by $\mathcal{V}$, only contains the initial point. Then, the LoD is generated by the following steps iteratively.

\begin{figure}[h]
\centering
\includegraphics[width=7cm,height=5.5cm]{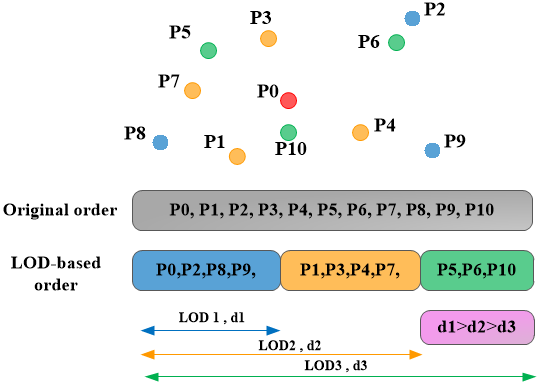}
\caption{LOD generation description. Red point P0 is the initial point.}
\end{figure}

\begin{enumerate}
\item Set $s$=1, ${dist_1}$=$MAX$, traverse all the points to construct the set of visited points $\mathcal{V}$ and the reorganized level set ${R_s}$: The current point would be neglected, if it has been visited; or else the minimum distance ${D_{\min }}$ of the current point to the points in the set $\mathcal{V}$ is calculated. If ${D_{\min }}$ is strictly smaller than ${dist_s}$, the current point is neglected, otherwise, the current point would be marked as visited and put into $\mathcal{V}$ and
    ${R_s}$.
\item The $s$-th LoD, denoted as LoD$_s$, is acquired by taking the union of the reorganized level sets ${R_1},...,{R_s}$.
\item  $s \leftarrow s + 1$, and go to 1), until all the LoDs are built or all the points have been marked as visited.
\end{enumerate}

\section{Test Model Categories for 3D Point Cloud Compression}
\vspace{3ex}
Since TMC1 and TMC3 have been merged into the new platform TMC13, in the following, we will first introduce the core technologies of TMC13, and then describe the core technologies of TMC2.

In TMC13, there are two kinds of encoder choice, i.e., the RAHT-based encoder (corresponds to the previous TMC1 [15]), and the LoD-based encoder (corresponds to the previous TMC3 [17]).
\vspace{-1ex}
\subsection{RAHT-based Encoder of TMC13 (TMC1, G-PCC [41])}
Chou, Nakagami, and Jang [15] first proposed RAHT-based point cloud compression platform, i.e., TMC1 in October 2017. Fig. 4 shows the encoder architecture of TMC1. As we can see that geometry compression and color compression are separately processed. The “coordinates conversion” module converts original geometry (namely world coordinates) to normalized geometry (namely frame coordinates). The quantization module will down-sample the points with a predefined quantization scale. The geometry encoder and decoder modules use octree and triangulation to encode/decode geometry information. The decoded geometry points are then recolored by the re-color module. The re-colored geometry is then encoded by a “color encoder” module in which the RAHT is used to efficiently compress the color information.
\subsubsection{Coordinates Conversion}
The geometry information of a 3D point cloud is first normalized from the 3D world coordinate to the frame coordinates:
\begin{equation}
({x_i},{y_i},{z_i}) = ((x_i^{world},{y_i^{world}},z_i^{world}) - ({t_x},{t_y},{t_z}))/\alpha,
\end{equation} where $({t_x},{t_y},{t_z})$ and $\alpha$ are translation and scale parameters, $(x_i^{world},{y_i^{world}},z_i^{world})$ and $({x_i},{y_i},{z_i})$ are the 3D world coordinate and the corresponding frame coordinate of the point $i$, $i \in 1,...,N$.

\begin{figure}[h]
\centering
\includegraphics[width=9cm,height=6cm]{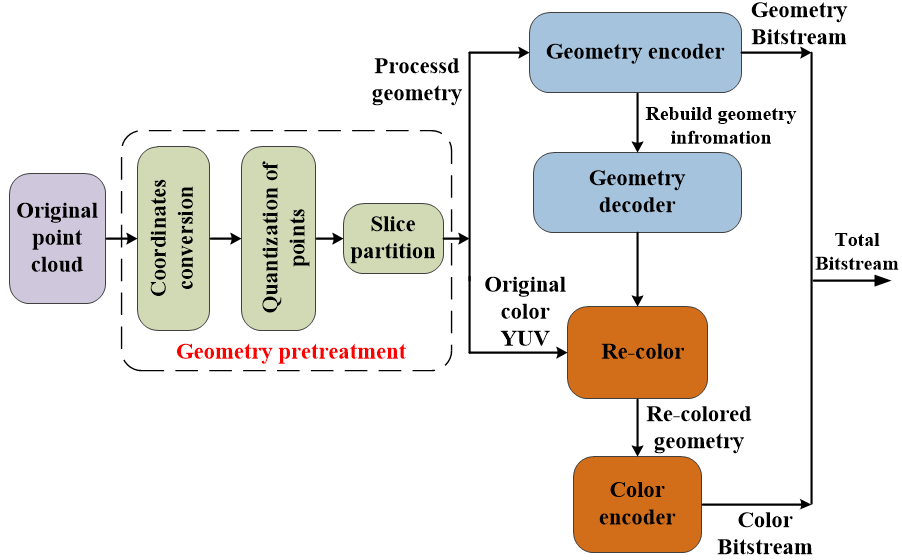}
\caption{TMC1 encoder architecture.}
\end{figure}
\vspace{-2ex}

\subsubsection{Quantization of Points}
After coordinate’s conversion, the coordinates of a 3D point cloud can be quantized depending on user requirement.
 Let ${\bm{X}_i} =
({x_i},{y_i},{z_i})$ denote the 3D coordinate of a point. The
quantized coordinates
$ {\bm{\mathord{\buildrel{\lower3pt\hbox{$\scriptscriptstyle\smile$}}
\over X}} _i} = ({\mathord{\buildrel{\lower3pt\hbox{$\scriptscriptstyle\smile$}}
\over x} _i},{\mathord{\buildrel{\lower3pt\hbox{$\scriptscriptstyle\smile$}}
\over y} _i},{\mathord{\buildrel{\lower3pt\hbox{$\scriptscriptstyle\smile$}}
\over z} _i}) $ can be calculated by

 \begin{equation}
 {\bm{\mathord{\buildrel{\lower3pt\hbox{$\scriptscriptstyle\smile$}} \over X}} _i} = Round(({\bm{X}_i} - {\bm{X}_{\min }}){\rm{ \times }}q),
 \end{equation} where $q$ is a position quantization scale factor, ${\bm{X}_{\min }} = ({x_{\min }},{y_{\min }},{z_{\min }})$, and ${x_{\min }}$, ${y_{\min }}$, and ${z_{\min }}$ are the
 minimum coordinates along the three axes of all the points. Note that there may exist points with the same coordinates after the coordinate quantization. The TMC13 platform also provides an option to remove the duplicated points.

\subsubsection{Slice Partition}
There are two slice partition methods in TMC13. One is the longest edge-based uniform partition. That is to say, the minimum and the maximum edge (${edge_{min}}$ and  ${edge_{max}}$) among the three directions \textit{x}, \textit{y}, and \textit{z}, are first calculated. Then, ${edge_{min}}$ is used as the slice partition interval, to cut the point cloud along with the direction whose edge is the longest. The remainder of points are also considered as slices. Another slice partition method is the octree-based uniform partition. In this method, each side of the bounding box of the point cloud is rounded up to a power of 2, and a pre-defined parameter ${depth_{partition}}$ is used as the termination condition of the octree decomposition, as described in Part A of Section III.

\begin{figure*}[!htb]
\centering
\includegraphics[width=14.5cm,height=5.1cm]{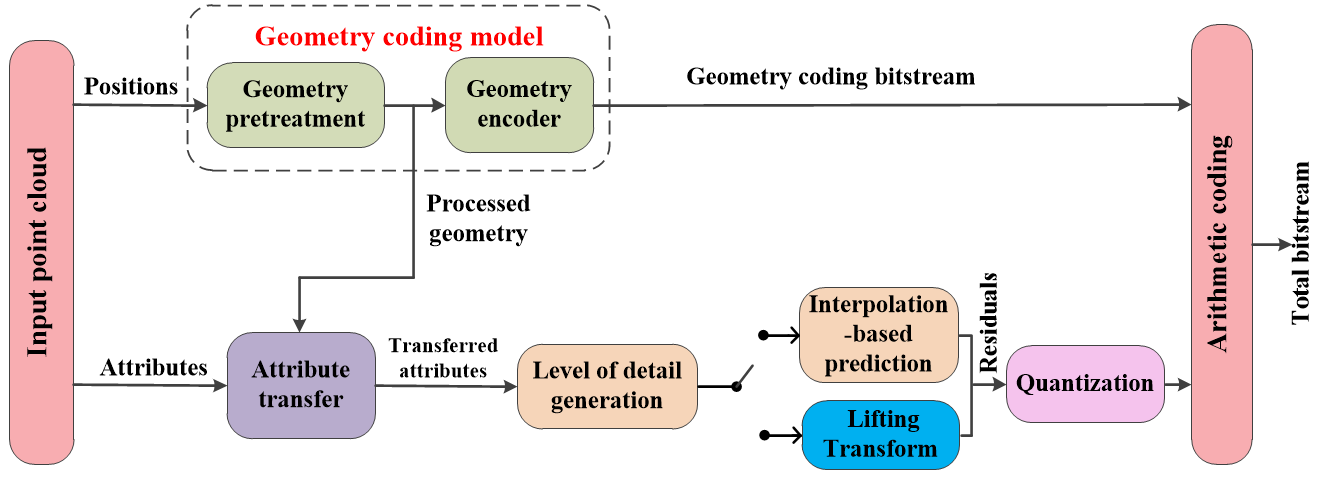}
\caption{ TMC3 encoder architecture.}
\end{figure*}

\subsubsection{Geometry Encoder}
In the geometry encoder module, octree-based decomposition is further conducted on all the partitioned slices. A bounding box is first constructed for each slice. The partitioned slices can then be divided into voxels with size of $W{\rm{ \times }}W{\rm{ \times }}W$, where $W = {2^{d - l}}$, \textit{d} denotes the octree depth, and \textit{l} denotes the octree level that corresponds to the available octree depth which is less than or equal to \textit{d}. If \textit{d}=\textit{l}, there only exists one point in a voxel. In this case, the geometry information will be compressed losslessly; otherwise, it will be compressed lossily.

\subsubsection{Lossless Geometry Compression}
In this case, the point cloud will be octree decomposed finely until there exists only one point in a voxel, i.e., \textit{d}=\textit{l}, $W$=1. If the points are distributed uniformly, the geometry information can be efficiently compressed by recoding the octree decomposition procedure. However, when there are isolated points, the isolated points should also be decomposed repeatedly to record their fine positions in the bounding box. Consequently, a lot of unnecessary bits must be consumed. In order to avoid this circumstance, a technique namely direct coding mode (DCM) [42] is proposed. In this method, the coordinates of the isolated points are extracted and encoded independently.
Furthermore, a neighbor-dependent entropy context [43] is used for entropy coding. In this method, to calculate the probability distribution of semantic symbols for the successive arithmetic encoder, 10 context patterns are proposed and updated separately based on the neighboring structure of the current node. In order to further improve the coding efficiency, an occupancy prediction method is also proposed in [44] for the context update.
\subsubsection{Lossy Geometry Compression}
In this case, \textit{l} is smaller than \textit{d}, i.e., the geometry loss is inevitable. The geometry information within a voxel with the size of $W{\rm{ \times }}W{\rm{ \times }}W$ is represented as a surface that intersects each edge of the voxel. Since there are 12 edges bounding on a voxel, there are at most 12 intersections between the surface and the voxel. The intersections are also called as vertices. A vertex along an edge can be discovered when there is no less than one occupied voxel adjacent to the edge among the whole blocks that share the edge. The location of a discovered vertex along corresponding edge is the average location along the edge of all such voxels adjacent to the edge among whole blocks that share the edge.

The vertices are then encoded as follows. Firstly, a set of flags are used to label whether an edge of a cube contains a vertex or not. Secondly, for the edge that includes a vertex, the relative location of the vertex on the edge is uniformly quantized. Finally, the set of flags and the quantized relative locations are entropy coded. By using these vertices, non-planar polygons (triangles) could be constructed in the geometry decoder, and the lost voxels could be approximately estimated by extracting the refined vertices from the constructed triangles.

Besides the triangle surface approximation-based lossy compression method, geometry information can also be quantized directly (2 of Part A of Section IV) for lossy compression namely direct geometry quantization method.
\subsubsection{Geometry Decoder}
Before compressing the color information, the geometry information must be decoded first. For losslessly compressed geometry, direct entropy decoder is enough, while for lossy compressed geometry, surface reconstruction will be performed after entropy decoder of the vertices. Then, non-planar triangles will be constructed based on the decoded vertices. Finally, refined vertices $({{\tilde x}_j},{{\tilde y}_j},{{\tilde z}_j})$, $j \in 1,...,M$, could be obtained by up-sampling these triangles so as to approximately estimate the lost geometry information [41].
\subsubsection{Re-Color}
The re-color module assigns original voxel colors to the refined vertices. It is implemented by assigning the color $({{\tilde
Y}_j},{{\tilde U}_j},{{\tilde V}_j})$ that is equal to $({Y_i},{U_i},{V_i})$ to a refined vertex $({{\tilde x}_j},{{\tilde y}_j},{{\tilde z}_j})$, where $i$ is the index of voxel position
$({x_i},{y_i},{z_i})$ that is closest to $({{\tilde x}_j},{{\tilde y}_j},{{\tilde z}_j})$.
\subsubsection{Color Encoder}
The reconstructed voxel colors $({{\tilde Y}_j},{{\tilde U}_j},{{\tilde V}_j})$ are then transformed by RAHT, uniformly quantized, and entropy coded. To see the detailed procedure of RAHT, please refer to [31].

\subsection{LOD-based Encoder of TMC13 (TMC3, G-PCC)}
LOD-based TMC13 (i.e., TMC3) is  proposed by \textit{Apple Inc}. in October 2017. As shown in Fig. 5, geometry information of a 3D point cloud is also compressed by octree decomposition, which is the same with the RAHT-based TMC13 encoder (i.e., TMC1). The color information is then differentially encoded by a LoD generation procedure.

\subsubsection{Geometry Pretreatment {\rm{\&}} Geometry Encoder}
The geometry preprocessing and geometry encoder are the same with the RAHT-based TMC13, as shown in Part A of current section.
\subsubsection{Attributes Transfer}
The color information of the original 3D point cloud is then transferred to the quantized coordinates by k-d tree to find the minimum Euclidian difference between the original coordinates and the reconstructed coordinates generated from inverse quantization.
\subsubsection{LoD Generation}
Afterwards, the geometry information of the original point cloud is re-organized by LoD, as shown in the part C of Section III. Besides, there are some other similar LOD generation methods i.e., scalable complexity-based method [45] and binary tree-based method [41], that are also adopted in TMC13.

\subsubsection{Interpolation-based Prediction (Predict Transform)}
The color information is then predicted and differentially encoded in the order defined by the LoD generation procedure. Note that only the already encoded/decoded points could be taken as reference points for prediction. For the $i$-th point, its attribute value ${c_i}$ can be predicted by the linear combination of its $k$ nearest neighbors:

\begin{equation}
{\hat c_i} = Round{\left(\sum\nolimits_{j \in {\Im _i}}
{\left(\frac{{\frac{1}{{\eta _j^2}}}}{{\sum\nolimits_{j \in {\Im_i}} {\frac{1}{{\eta _j^2}}} }}\right)}{\tilde c_j}\right)},
\end{equation} where ${\Im _i}$ is the set of the $k$-nearest neighbors that are already decoded of the $i$-th point, ${{\tilde c}_j}$ is the corresponding reconstructed attribute values, ${\eta _j}$ is the distance between the $i$-th point and the $j$-th neighbor. Finally, residual ${r_i}$ between the attribute value of the $i$-th point ${c_i}$ and the predicted attribute value ${\hat c_i}$ is quantized and arithmetically encoded.


\begin{figure}
\centering
\includegraphics[width=8.8cm,height=4.7cm]{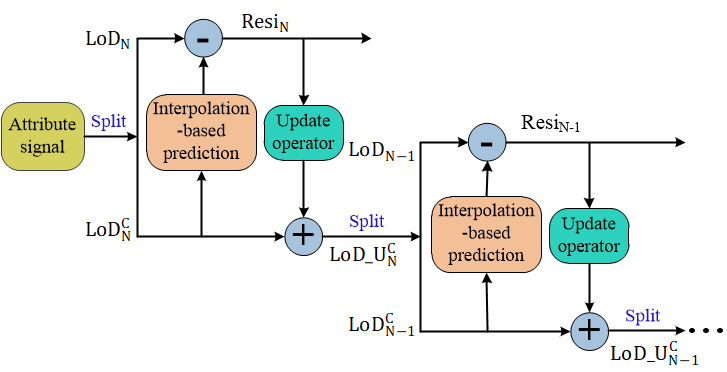}
\caption{Lifting transform scheme.}
\end{figure}
\subsubsection{Lifting Transform (LT)}
To further improve the coding efficiency, a LT is proposed in [41]. The basis of this method is to make the points in the lower LODs [45] more influential (i.e., because they are frequently used to make predictions). In this method, the points (denoted as $\bm{\Psi}$) in lower LoD that are used to predict the points (denoted as $\bm{\Phi}$) in the $N$-th LoD is updated by the residual of the points in the $N$-th LoD, as shown in Fig. 6. In this figure, the attribute values of points in the $N$-th LoD (denoted as LoD$_N$) are predicted by the attribute values in the other LoDs (denoted as LoD$_N^C$) whose LoD order is smaller than $N$. Let ${Resi_{N}}$ denotes the predicted attribute residuals of the points in the $N$-th LoD, LoD\_U$_N^C$ represents the updated attributes in lower LoDs, where $N$ is largest LOD order. The update procedure is described as follows.

Take the points in the $N$-th LoD as an example. Let $\psi$ be a point in $\bm{\Psi}$. Then, find the point set (i.e., $\bm{\Phi}$ in the $N$-th LoD) that are partially predicted (influenced) by $\bm{\Psi}$. The attribute values of updated point $Att(\psi')$ in $\bm{\Psi}$ can be calculated by

\begin{equation}
Att(\psi') = Att(\psi) + \frac{{\sum\nolimits_i^K {{\zeta _i} \times w({\phi _i}) \times  Resi(i)} }}{{\sum\nolimits_i^K {{\zeta _i} \times w({\phi _i})} }},
\end{equation} where $Att(\psi)$ is the attribute value of point $\psi$, ${\phi_i}$ is a point in $\bm{\Phi}$, $w({\phi_i})$ is the corresponding influence weight (with initial value of 1), \textit{K} denotes the number of points that are influenced by ${\psi}$, ${\zeta_i}$ depends on the Euclidean distance between by ${\psi}$ and ${\phi_i}$. Afterwards, the influence weight $w({\psi})$ of the point ${\psi}$ in $\bm{\Psi}$ can be calculated by

\begin{equation}
w(\psi ) = w(\psi ) + \sum\nolimits_i^K {{\zeta _i}{\rm{ \times }}w({\phi _i})}.
\end{equation}

Similarly, the updated points in the $(N-1)$-th LoD can be processed based on the above procedure recursively, until the last LoD is reached.

The residuals are finally quantized and entropy encoded to generate a bit stream. During the quantization procedure, an additional technique namely adaptive quantization [41] was also adopted in the LoD-based TMC13. In the adaptive quantization, the influence weights of each point are used to guide the quantization, i.e., the coding attribute residual of a point is further multiplied by the root of the influence weight of the point before the quantization. Note that it is no need to encode the influence weights of each points because they can be calculated during the lifting transform iteratively.

\begin{figure*}
\centering
\includegraphics[width=16cm,height=7.3cm]{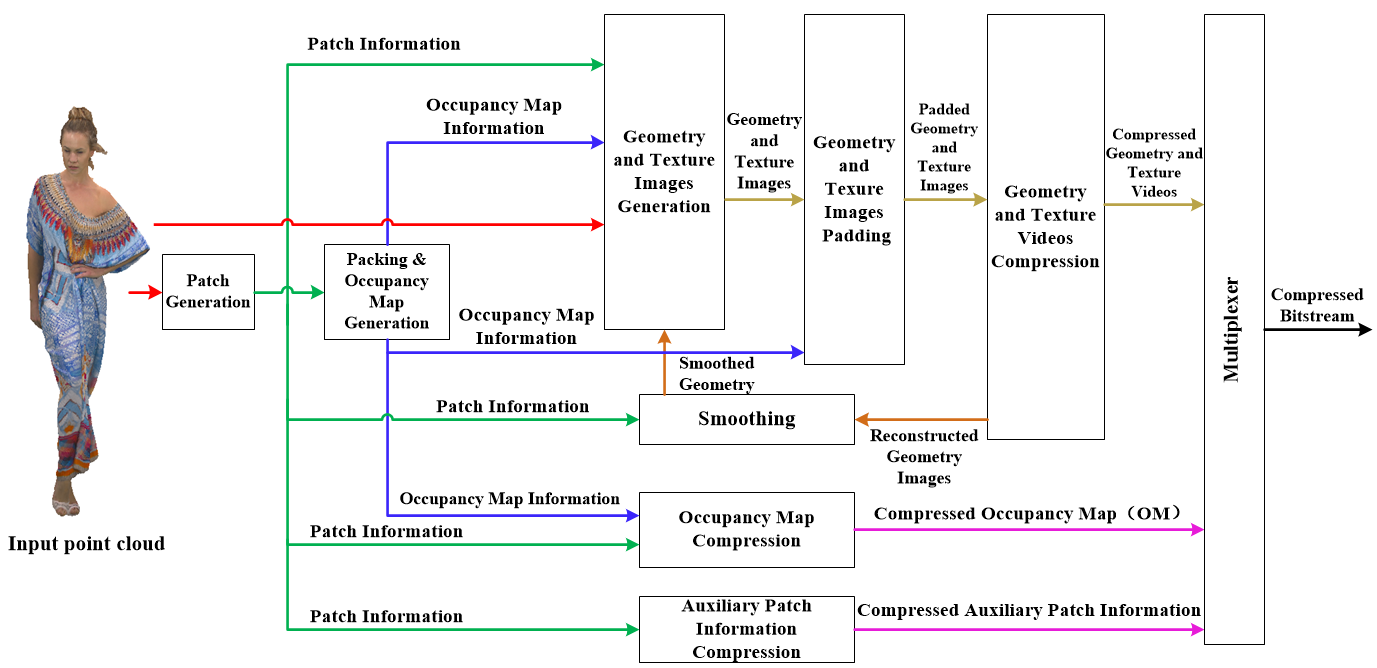}
\caption{ TMC2 encoder architecture.}
\end{figure*}

\subsection{Test Model Category 2 (TMC2, V-PCC [46])}
TMC2 was also proposed by ${Apple}$ ${Inc}$. in October 2017, the main idea of TMC2 is to convert the point cloud data into a set of video sequences and use state-of-the-art video codec (e.g., HEVC) to compress the geometry and texture information, as shown in Fig. 7.

\subsubsection{Patch Generation}
Given a 3D point cloud, the normal of each point is first computed by principal component analysis (PCA) based on the set of neighboring points. By utilizing normal information, an initial clustering of the point cloud is obtained through mapping each point to six predefined planes. The mapping operation could be performed by maximizing the dot product of the point normal and the pre-defined plane normal. Then, the spatially neighboring pixels in a cluster are extracted as patches.
\subsubsection{Packing {\rm{\&}} Occupancy map generation}
Afterwards, the extracted patches are mapped onto a 2D grid (with size of $W{\rm{ \times }}H$) by a packing procedure during which the occupancy information of each of the 2D grid is also recorded. The packing procedure aims at mapping the extracted patches onto a 2D grid, while trying to minimize the unused space of the 2D grid and guarantee that the pixels in each block (e.g., with size of $16{\rm{ \times }}16$) of the 2D grid corresponds to a unique patch. The occupancy map (OM) is a binary map that indicates each block of the 2D grid whether it belongs to the point cloud or empty. Specifically, each block consists of multiple smaller $h{\rm{ \times }}h$ (e.g., $h$ = 4) sub-blocks. If a sub-block is marked as 1 (namely full), then it must include at least a real (non-padded) pixel. Otherwise it is marked as 0 (namely non-full). When all sub-blocks in a block are marked 1, then corresponding block is labeled 1, and 0 otherwise. Note that above all marked binary flags would be encoded as additional information [46]. Fig. 8 (a) shows a block-based occupancy map.

\begin{figure}
\centering
\includegraphics[width=9cm,height=4.4cm]{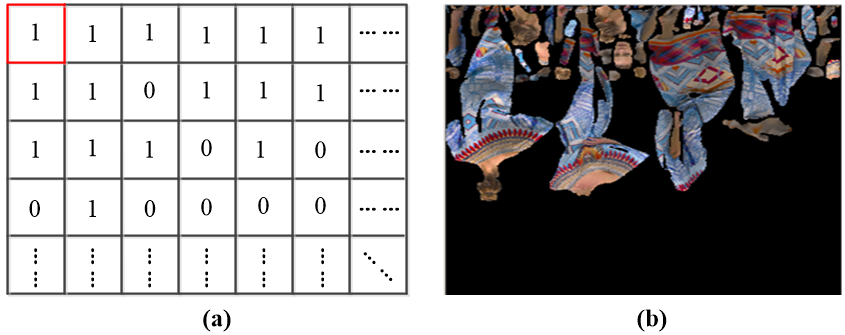}
\caption{(a) A block-based occupancy map. Occupancy map with size of $W{\rm{ \times }}H$, a block (red block in upper left corner) with size 16${\rm{ \times }}$16 was indicated 1, if all sub-blocks of it are marked as full, and 0 otherwise. (b) A mapped 2D point cloud texture image [46] is generated by image generation (the empty pixels are filled in black).}
\end{figure}

\subsubsection{Image Generation}
3D to 2D mapping is exploited to convert the geometry and texture of the point cloud into images, as shown in Fig. 8 (b). To overcome the problem that several points may be projected onto the same pixel, each patch is projected onto two images called near layer and far layer. More precisely, let $I(u,v)$ denote the set of points in the current patch that would be projected to the same pixel position $(u,v)$. The near layer will store the point of $I(u,v)$ with the minimum depth ${D_0}$, while the far layer, records the point with the maximum depth ${D_0} + \Delta$, where $\Delta$ refers to surface thickness parameter.
\subsubsection{Image Padding {\rm{\&}} Coding}
To generate piecewise smooth images suited for video compression, the empty space between patches in images is filled with the neighbors by a padding procedure. This smoothing operation is designed to alleviate potential compression artifacts caused by the discontinuities at the patch boundaries. Note that the padding points are not counted as occupied point. Finally, the generated occupancy map, geometry and texture images are encoded by HEVC as video frames. Different quantization steps in the HEVC encoding of the geometry and color video frames give different distortions and bit rates.

It is worthy pointing out that the related technologies are still ongoing. Recently, several new technologies, i.e., color and geometry smoothing [47], patch flexible orientation [48], push-pull background filling [49], interleaved absolute depth and color coding [50], patch-based color sub-sampling [51], and spatially adaptive geometry interpolation [52], were proposed and implemented in the newest version of TMC2 by MPEG PCC group, to further improve its performance.

\section{Experimental Results and Analysis}

\begin{figure*}[!htb]
\centering
\includegraphics[width=13cm,height=9.8cm]{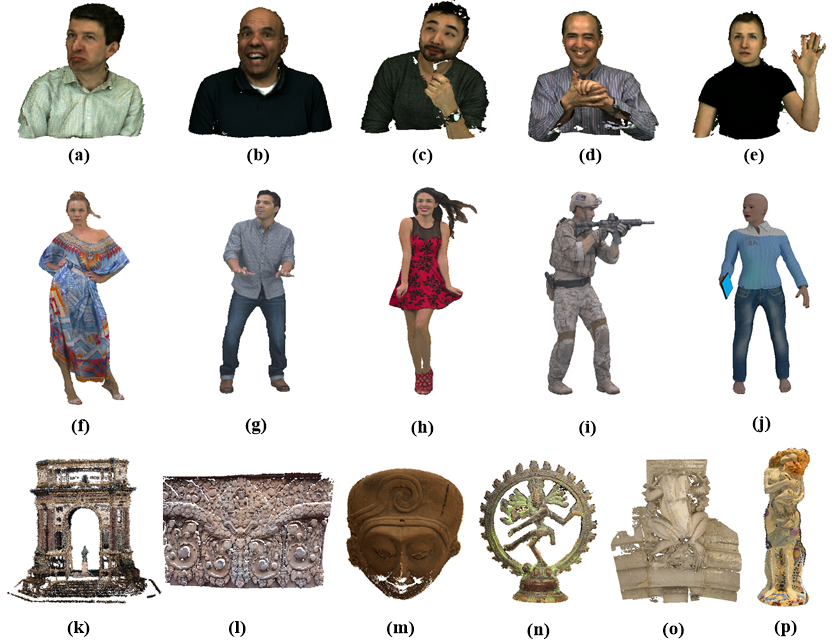}
\caption{3D point cloud sequences used in experiments. (a) Andrew, (b) Ricardo, (c) David, (d) Phil, (e) Sarah, (f) Longdress\_vox10\_1300, (g) Loot\_vox10\_1200, (h) Redandblack\_vox10\_1550, (i) Soldier\_vox10\_0690, (j) Queen\_0200, (k) Arco\_Valentino\_Dense\_vox12, (l) Facade\_00009\_vox12, (m) Egyptian\_mask\_vox12, (n) Shiva\_00035\_vox12, (o) Frog\_00067\_vox12, (p) Staue\_Klimt\_vox12.}
\end{figure*}

To verify the performance of the test model categories, extensive experiments were conducted over 16 3D point cloud datasets, as shown in Fig. 9, in which the sub-figures (a)-(j) are portraits with dense points, and (k)-(p) are engravings with sparse points. Sub-figures (a)-(e) are chosen from dynamic voxelized point cloud data sequences of Microsoft Voxelized Upper Bodies [53], and the corresponding extracted frame indexes are 0, 39, 0, 1, 18, respectively. Sub-figures (f)-(j) were obtained from 8i Voxelized Full Bodies [54]. Remaining large-scale sparse point clouds in sub-figures (k)-(p) were extracted from static objects and scenes datasets of test class A in common test conditions (CTC) [55] that can be downloaded from [56]. In the experiments, the RGB formatted attributes were first converted to the YUV format for compression. The reference software [57], TMC13 version 5 (TMC13 V5) and TMC2 version 5 (TMC2 V5) were used.
To evaluate the quality of reconstructed 3D point clouds, point to point distortion metric [58] was used for both geometry and color information. Based on the point to point distortion, the corresponding PSNRs of geometry and color information were then calculated by

\begin{equation}
\setlength{\abovedisplayskip}{0.7pt}
\setlength{\belowdisplayskip}{0.7pt}
{PSN{R_g}} = 10lo{g_{10}}({\Omega^2}/{d_{s\_rms}}{({{\bm{P}}_{ori,g}},{{\bm{P}}_{rec,g}})^2}),
\end{equation}

\begin{equation}
\setlength{\abovedisplayskip}{0.7pt}
\setlength{\belowdisplayskip}{0.7pt}
{PSN{R_c}} = 10lo{g_{10}}({255^2}/{d_{s\_rms}}{({{\bm{P}}_{ori,c}},{{\bm{P}}_{rec,c}})^2}),
\end{equation} where

\begin{equation}
\setlength{\abovedisplayskip}{0.7pt}
\setlength{\belowdisplayskip}{0.7pt}
\begin{aligned}
{d_{s\_rms}}({{\bm{P}}_{ori,\varphi}}{\rm{,}}{{\bm{P}}_{rec,\varphi}}) =
 max({d_{rms}}({{\bm{P}}_{ori,\varphi }}{\rm{,}}{{\bm{P}}_{rec,\varphi }}),
 \\ {d_{rms}}({{\bm{P}}_{rec,\varphi}}{\rm{,
}}{{\bm{P}}_{ori,\varphi}})),
\end{aligned}
\end{equation}

\begin{equation}
\setlength{\abovedisplayskip}{0.7pt}
\setlength{\belowdisplayskip}{0.7pt}
{d_{rms}}({\bm{P}_A},{\bm{P}_B}) = \sqrt {\frac{1}{k}\sum\nolimits_{{\bm{p}}
\in {{\bm{P}}_{A}}} {\|{{{\bm{p}}_\varphi } -
{{\bm{p}}_{B\_nn,\varphi }}}\|_2^2}},
\end{equation}

\begin{equation}
\setlength{\abovedisplayskip}{0.7pt}
\setlength{\belowdisplayskip}{0.7pt}
\begin{aligned}
\Omega  = max(({x_{max}} - {x_{min}}),({y_{max}} - {y_{min}}),\\
                                     ({z_{max}} - {z_{min}})),
\end{aligned}
\end{equation} ${{\bm{P}}_{ori}}$ and ${{\bm{P}}_{rec}}$ are original and reconstructed point clouds, the subscript $\varphi  \in \{ g,c\}$ denotes the geometry or  color information, ${{\bm{P}}_{A}}$ and ${{\bm{P}_{B}}}$ are two arbitrary point clouds that corresponds to ${{\bm{P}}_{ori,\varphi}}$ and ${{\bm{P}}_{rec,\varphi}}$, $k$ is the number of points in ${{\bm{P}}_{A}}$ , ${\bm{p}}$ denotes a point in ${{\bm{P}}_{A}}$, ${{\bm{P}}_{B\_nn}}$ is the nearest neighbor of ${\bm{p}}$ in ${{\bm{P}_{B}}}$, ${x_{max}}$ , ${x_{min}}$ , ${y_{max}}$ , ${y_{min}}$ , ${z_{max}}$ , ${z_{min}}$ are the maximum and the minimum coordinates along with the $x$, $y$, and $z$ axis. Similar to traditional image/video compression, the rate-distortion curves and the BD-PSNR (resp. BD-BR) [59] were used to do qualitative and quantitative comparisons respectively.
The performance of TMC13 and TMC2 are compared under 3 cases:\\
\textbf{\emph{Case 1, both geometry and color are compressed losslessly}};\\
\textbf{\emph{Case 2, geometry is compressed losslessly, while the color is compressed lossily}};\\
\textbf{\emph{Case 3, both geometry and color are compressed lossily}}.\\

\vspace{-1.5ex}

To conduct the experiments, a computer equipped with Intel Core i7 8700K CPU (3.7GHz frequency), 64GB memory, and windows 10 operating system was used. The main software configurations of the three cases are shown in Table I. Note that LODC is the level of detail count, PQS indicates position quantization scale, DBODL defines the difference between octree depth and level, which can be used for triangle surface approximate. RQS represents RAHT color quantization steps (both luma and chroma), LTQS refers to quantization steps (both luma and chroma) of LT, and PTQS means color quantization steps (both luma and chroma) of predict transform (refers to the LoD-based encoder without LT). GQP expresses geometry quantization steps, while TQP denotes the texture (color) quantization step of TMC2. Other parameters were set according to the parameter encoding files provided in the TMC13 and TMC2 software packages.

\begin{table*}[htbp]
  \centering
  \caption{Main Coding Parameters for TMC13 and TMC2.}
  \renewcommand\tabcolsep{6pt} 
    \begin{tabular}{cccp{4.44em}cccccc}
    \toprule
    \toprule
    \textbf{Coding} & \multicolumn{6}{c}{\textbf{TMC13}} &    & \multicolumn{2}{c}{\textbf{TMC2}} \\
\cmidrule{2-7}\cmidrule{9-10}    \textbf{parameters} & \textbf{LODC} & \textbf{PQS} & \multicolumn{1}{c}{\textbf{DBODL}} & \textbf{RQS} & \textbf{LTQS} & \textbf{PTQS} &    & \textbf{GQP} & \textbf{CQP} \\
\cmidrule{1-7}\cmidrule{9-10}    \textbf{Case1} & 8  & 1  & \multicolumn{1}{c}{---} & --- & --- & 0  &    & --- & --- \\
    \textbf{Case2} & 8  & 1  & \multicolumn{1}{c}{---} & 2 - 26 & 2 - 26 & 2 - 32 &    & --- & --- \\
    \textbf{Case3} & 8  & 0.9, 0.7, 0.4, 0.2 & 1, 2, 3, 4 & 2 - 32 & 2 - 32 & 2 - 32 &    & 6 - 40 & 2 - 40 \\
    \bottomrule
    \bottomrule
    \end{tabular}%
  \label{tab:addlabel}%
\end{table*}%

\begin{table*}[htbp]
  \centering
  \caption{Rate and Coding Time Comparisons of TMC13 and TMC2 for Case 1.}
  \renewcommand\arraystretch{0.9}
   \renewcommand\tabcolsep{3pt} 
    \begin{tabular}{ccccccccccc}
    \toprule
    \toprule
       & \multicolumn{4}{c}{\textbf{TMC13 }} &    & \multicolumn{4}{c}{\textbf{TMC2}}  \\
\cmidrule{2-5}\cmidrule{7-11}    \textbf{Datasets} & \textbf{Geometry} & \textbf{Color} & \textbf{Total} & \textbf{Coding} &    & \textbf{Occupancy} & \textbf{Geometry} & \textbf{Color} & \textbf{Total} & \textbf{Coding} \\
       & \textbf{rate (bpp)} & \textbf{rate (bpp)} & \textbf{rate (bpp)} & \textbf{Time (s)} &    & \textbf{map rate (bpp)} & \textbf{rate (bpp)} & \textbf{rate (bpp)} & \textbf{rate (bpp)} & \textbf{Time (s)} \\
    \midrule
    \textbf{Andrew} & 1.17  & 15.19  & 16.36  & 2.30 &    & 0.18  & 1.56 & 13.99 & 15.74 & 50.98 \\
    \rowcolor[rgb]{ .816,  .808,  .808} \textbf{Ricardo} & 1.08  & 8.13  & 9.22  & 1.66 &    & 0.14  & 1.36 & 8.04 & 9.55 & 38.24 \\
    \textbf{David} & 1.14  & 10.13  & 11.27  & 2.64 &    & 0.14  & 1.47 & 10.11 & 11.73 & 58.09 \\
    \rowcolor[rgb]{ .816,  .808,  .808} \textbf{Phil} & 1.21  & 14.12  & 15.33  & 3.03 &    & 0.22  & 1.66 & 13.73 & 15.61 & 67.40 \\
    \textbf{Sarah} & 1.14  & 6.88  & 8.02  & 2.40 &    & 0.19  & 1.53 & 6.92 & 8.64 & 51.16 \\
    \rowcolor[rgb]{ .816,  .808,  .808} \textbf{Longdress\_vox10\_1300} & 1.05  & 14.08  & 15.12  & 7.10 &    & 0.09  & 1.36 & 13.75 & 15.20 & 177.87 \\
    \textbf{Loot\_vox10\_1200} & 1.00  & 8.89  & 9.89  & 6.44 &    & 0.10  & 1.30 & 7.90 & 9.30 & 159.74 \\
    \rowcolor[rgb]{ .816,  .808,  .808} \textbf{Redandblack\_vox10\_1550} & 1.13  & 10.27  & 11.40  & 6.13 &    & 0.16  & 1.55 & 11.54 & 13.25 & 161.09 \\
    \textbf{Soldier\_vox10\_0690} & 1.06  & 11.09  & 12.15  & 8.73 &    & 0.10  & 1.43 & 8.91 & 10.45 & 219.87 \\
    \rowcolor[rgb]{ .816,  .808,  .808} \textbf{Queen\_0200} & 0.80  & 9.72  & 10.51  & 8.00 &    & 0.13  & 0.99 & 8.72 & 9.86 & 186.66 \\
    \textbf{Arco\_Valentino\_Dense\_vox12} & 9.86  & 21.05  & 30.92  & 14.91  &    & 6.17  & 31.34  & 29.62  & 67.13  & 33309.60  \\
    \rowcolor[rgb]{ .816,  .808,  .808} \textbf{Facade\_00009\_vox12} & 7.25  & 14.97  & 22.23  & 15.83  &    & 3.69  & 13.31  & 27.09  & 44.09  & 13938.20  \\
    \textbf{Egyptian\_mask\_vox12} & 12.58  & 11.31  & 23.89  & 2.80  &    & 9.93  & 43.14  & 34.14  & 87.21  & 4598.19  \\
    \rowcolor[rgb]{ .816,  .808,  .808} \textbf{Shiva\_00035\_vox12} & 9.68  & 19.07  & 28.75  & 10.37  &    & 6.77  & 28.99  & 31.16  & 66.92  & 33465.60  \\
    \textbf{Frog\_00067\_vox12} & 6.44  & 13.08  & 19.52  & 35.07  &    & 3.74  & 10.84  & 26.53  & 41.11  & 21472.60  \\
    \rowcolor[rgb]{ .816,  .808,  .808} \textbf{Staue\_Klimt\_vox12} & 9.94  & 15.71  & 25.65  & 5.07  &    & 7.14  & 31.09  & 31.11  & 69.34  & 11143.30  \\
    \midrule
    \textbf{Average} & \textbf{4.16} & \textbf{12.73} & \textbf{16.89} & \textbf{8.28} &    & \textbf{2.43} & \textbf{10.81} & \textbf{17.70} & \textbf{30.95} & \textbf{7443.66} \\
    \bottomrule
    \bottomrule
    \end{tabular}%
  \label{tab:addlabel}%
\end{table*}%

\subsection{Results of Case 1}
In this case, the LoD-based encoder without LT (denoted as \textbf{LoD encoder w/o LT}) of TMC13 was used. Because the TMC2 V5 has some bugs in lossless compression for large-scale sparse point cloud (i.e., Fig. 9 (k)-(p)), TMC2 V6 was used to perform lossless compression for these point clouds. The corresponding coding bits per point (\emph{bpp}) of geometry and color are given in Table II. In this case, the \textbf{LoD encoder w/o LT} in TMC13 was used for the lossless compression of color information. We can see that, for the geometry information, the \emph{bpp} of TMC13 is lower than TMC2, i.e., the average geometry \emph{bpp} of TMC13 and TMC2 are 4.16 and 10.81 respectively. Similar results appear in the performance of color compression and coding complexity. The color \emph{bpp} and coding time of TMC13 is also lower than TMC2.

Accordingly, TMC13 is obviously more suitable for lossless compression.

\subsection{Results of Case 2}
Because the TMC2 cannot work in this case, we only compared the RAHT-based encoder (denoted as \textbf{RAHT encoder}), the \textbf{LoD encoder w/o LT} and the LoD based encoder with LT (denoted as \textbf{LoD encoder with LT}) of TMC13. Fig. 10 shows the rate-PSNR curves of color information for the three encoders. We can see that \textbf{LoD encoder with LT} is the best when the rate is low, while the \textbf{LoD encoder w/o LT} is usually the best when the rate is high. The performance of the \textbf{RAHT encoder} is in between with or lower than \textbf{LoD encoder with LT} and \textbf{LoD encoder w/o LT}, for different point clouds.

\begin{figure*}[htbp]
\centering
\includegraphics[width=15.5cm,height=11.2cm]{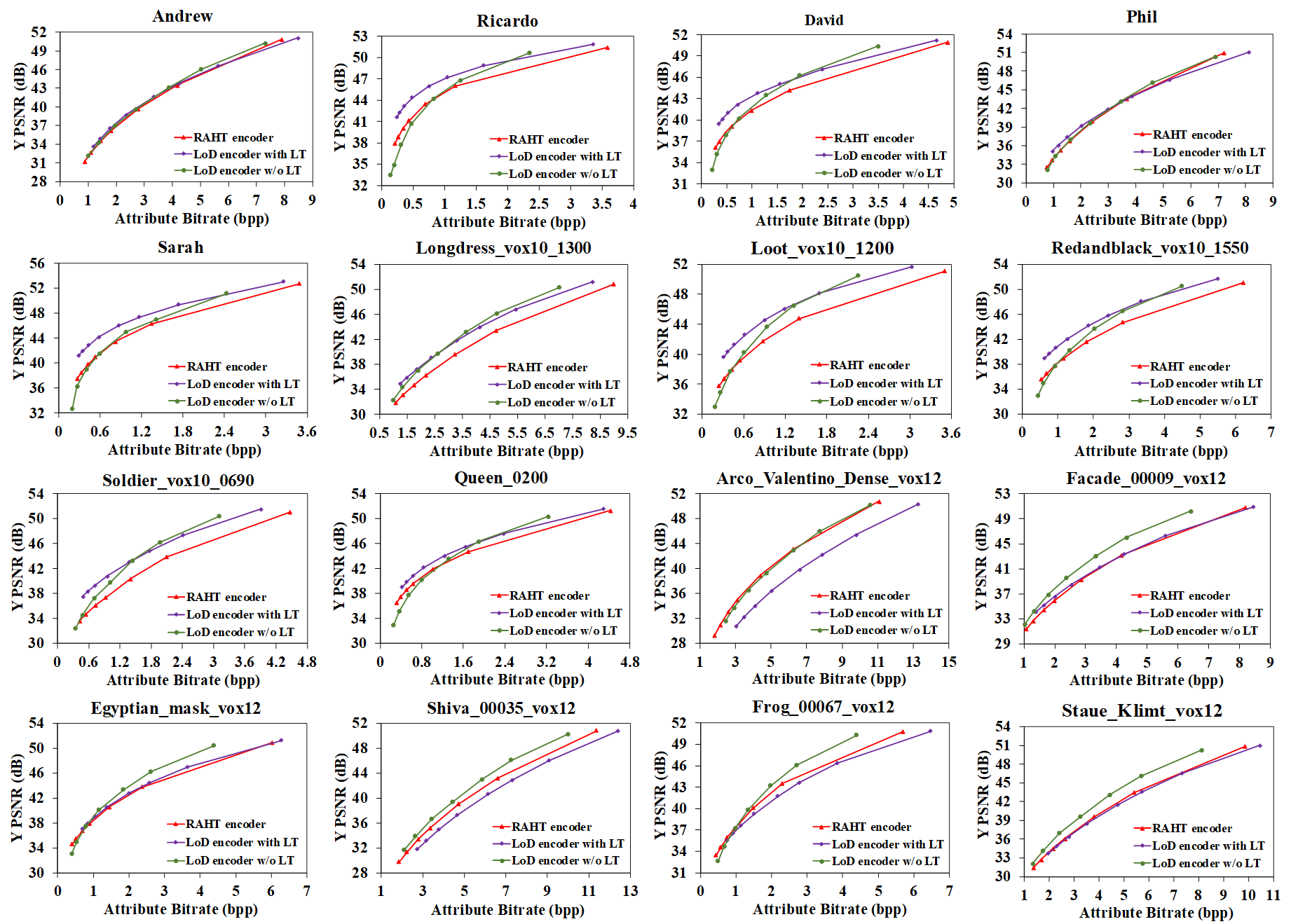}
\caption{Comparisons of rate-PSNR curves of case 2.}
\end{figure*}

\begin{figure*}[htbp]
\centering
\includegraphics[width=15.5cm,height=11.2cm]{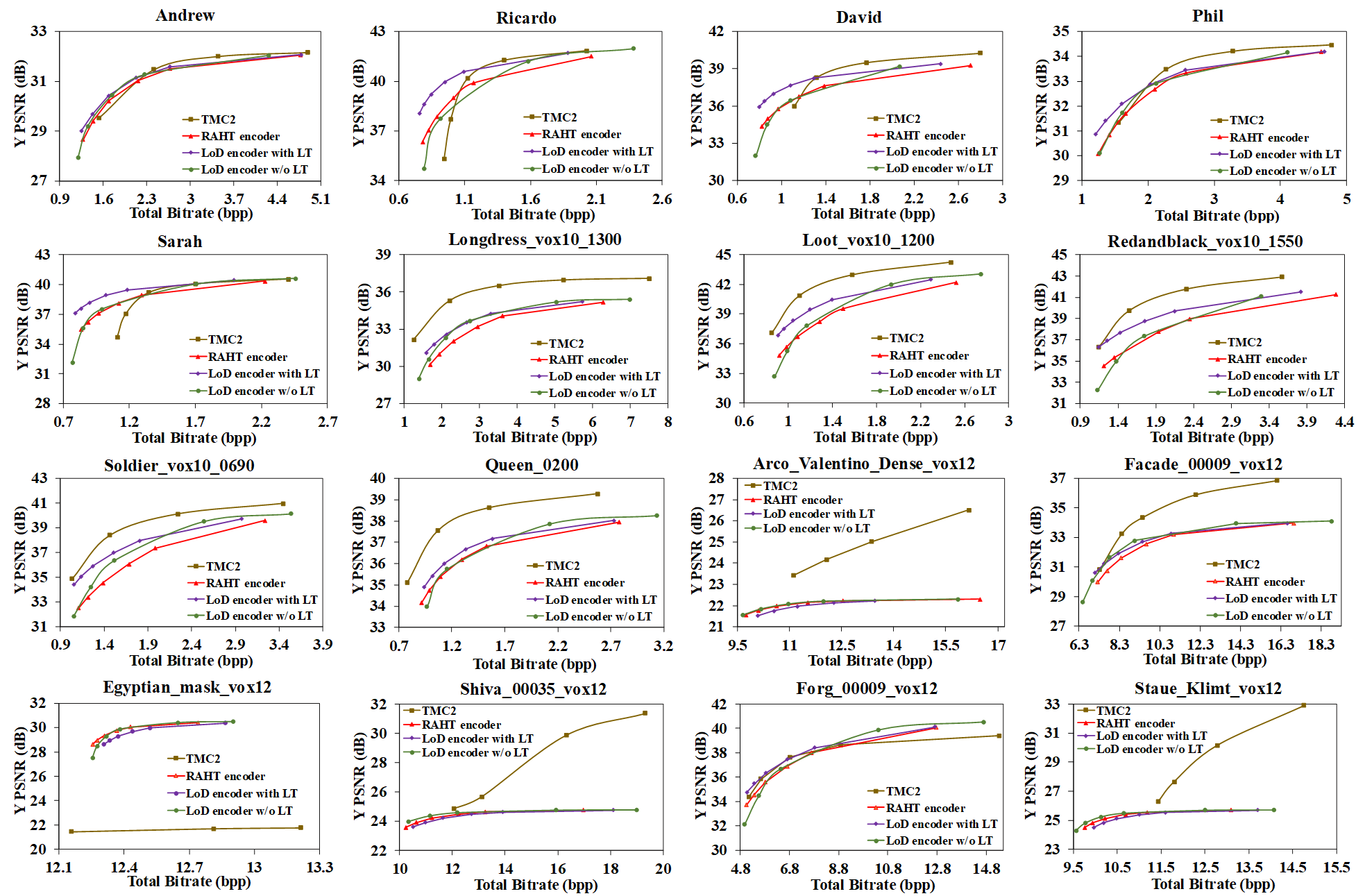}
\caption{Comparisons of rate-PSNR curves of case 3, in which the triangle surface approximation-based lossy geometry compression is used.}
\end{figure*}

Due to the limitation of the paper space, we put the detailed rate distortion data in the additional \textbf{Appendix}\footnote{The Appendix can be found in the additional appendix.}. The detailed comparisons of the three methods can be found in \textbf{Appendix A}, whereas Table III shows the BD-PSNR and BD-BR when using the \textbf{RAHT encoder} as the benchmark. From Table III, we can see that the performance of the Y component of \textbf{LoD encoder w/o LT} is better than other two methods, i.e., an average 0.84 dB BD-PSNR can be achieved by \textbf{LoD encoder w/o LT} when comparing it to \textbf{RAHT encoder}, while the corresponding BD-PSNR of the \textbf{LoD encoder with LT} is 0.77 dB higher than that of the \textbf{RAHT encoder}.
\begin{figure*}[htbp]
\centering
\includegraphics[width=13.9cm,height=14.1cm]{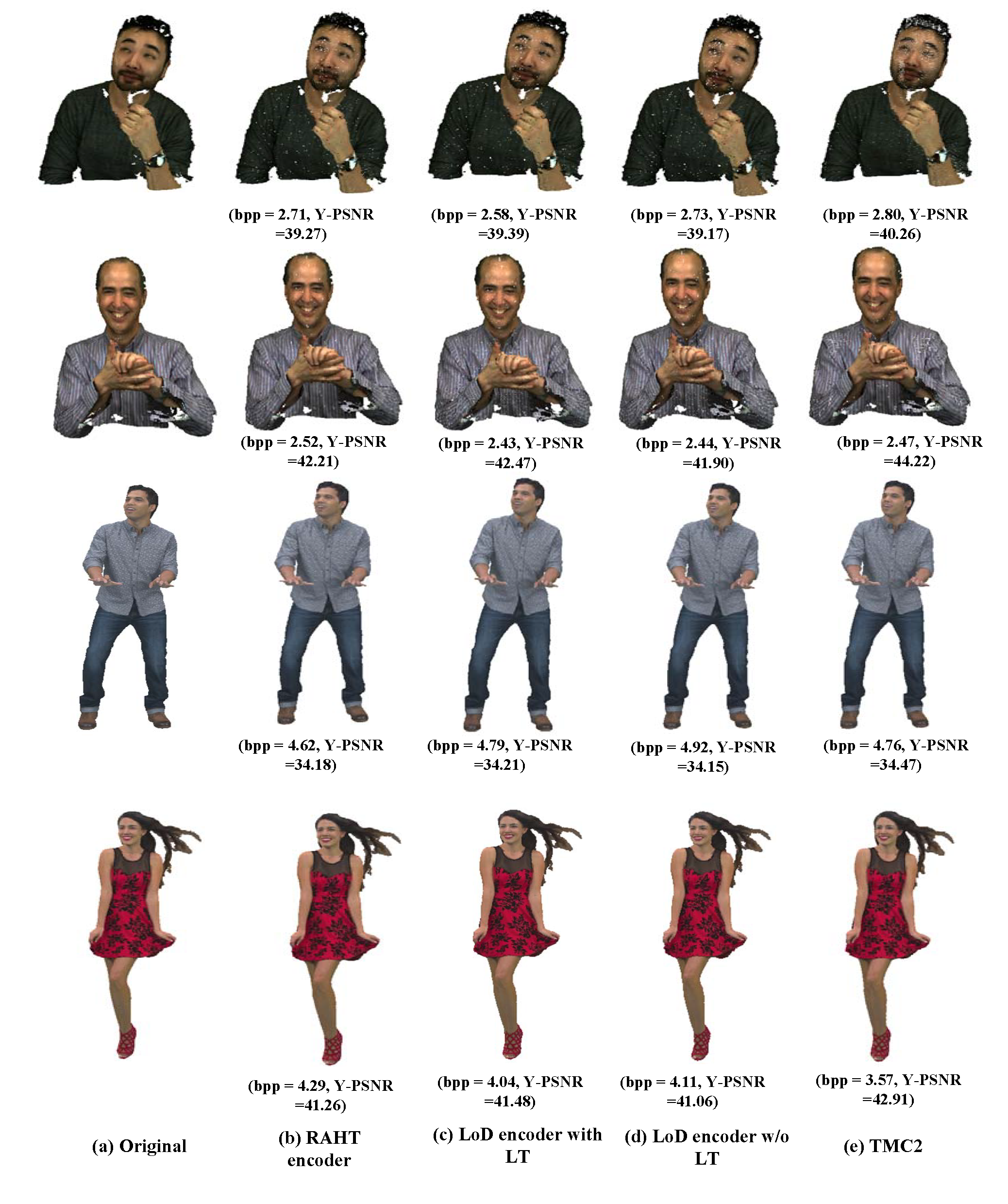}
\caption{Subjective quality comparisons of case 3, in which the triangle surface approximation-based lossy geometry compression is used.}
\end{figure*}

Besides, we also compared the performance of U and V components as shown in Table III. Because the local correlation of the U and V components are much higher than that of the Y component, the performance of the \textbf{LoD encoder with LT} is the best. We can also find that the performance of the \textbf{LoD encoder w/o LT} is the worst. Therefore, we can conclude that both the RAHT and the LT can efficiently utilize the correlation of color information, but the performance of LT is the better in this case.

\subsection{Results of Case 3}
In this case, for the TMC13, the geometry can be lossily compressed by two approaches, i.e., the triangle surface approximation-based method, and the direct geometry quantization-based method, as described in 2 of Part A of Section IV.

\begin{table*}[htbp]
  \centering
  \caption{BD-PSNR and BD-BR Comparisons of LoD Encoder w/o LT and LoD Encoder With LT for Case 2, RAHT Encoder is the Benchmark.}
  \renewcommand\tabcolsep{0.65pt} 
    \begin{tabular}{cccccccccccccc}
    \toprule
    \toprule
       & \multicolumn{6}{c}{\textbf{LoD encoder w/o  LT}} &    & \multicolumn{6}{c}{\textbf{LoD encoder with  LT}} \\
\cmidrule{2-7}\cmidrule{9-14}    \textbf{Datasets} & \textbf{Y-BD-} & \textbf{BD-BR} & \textbf{ U-BD-} & \textbf{BD-BR} & \textbf{V-BD-} & \textbf{BD-BR} &    & \textbf{Y-BD-} & \textbf{BD-BR} & \textbf{U-BD-} & \textbf{BD-BR} & \textbf{V-BD-} & \textbf{BD-BR} \\
       & \textbf{PSNR (dB)} & \textbf{(\%)} & \textbf{PSNR (dB)} & \textbf{(\%)} & \textbf{PSNR (dB)} & \textbf{(\%)} &    & \textbf{PSNR (dB)} & \textbf{(\%)} & \textbf{PSNR (dB)} & \textbf{(\%)} & \textbf{PSNR (dB)} & \textbf{(\%)} \\
\cmidrule{1-7}\cmidrule{9-14}    \textbf{Andrew} & \textcolor[rgb]{ 0,  .439,  .753}{\textbf{0.49 }} & \textcolor[rgb]{ 0,  .439,  .753}{\textbf{-4.62 }} & -1.18  & 17.85  & -1.24  & 18.40  &    & -0.05  & 0.65  & -0.50  & 7.86  & -0.55  & 7.92  \\
    \rowcolor[rgb]{ .816,  .808,  .808} \textbf{Ricardo} & -0.47  & 7.26  & -4.64  &   228.77$^*$  & -5.63  &   137.18$^*$ &    & \textcolor[rgb]{ 0,  .439,  .753}{\textbf{2.36 }} & \textcolor[rgb]{ 0,  .439,  .753}{\textbf{-40.44 }} & 1.14  & -45.57  & 1.17  & -34.96  \\
    \textbf{David} & 0.46  & -7.24  & -4.60  &   127.18$^*$  & -3.73  & 94.06  &    & \textcolor[rgb]{ 0,  .439,  .753}{\textbf{2.05 }} & \textcolor[rgb]{ 0,  .439,  .753}{\textbf{-37.29 }} & 1.09  & -38.14  & 1.03  & -34.72  \\
    \rowcolor[rgb]{ .816,  .808,  .808} \textbf{Phil} & 0.06  & -0.74  & -3.63  & 58.78  & -3.03  & 45.93  &    & \textcolor[rgb]{ 0,  .439,  .753}{\textbf{0.47 }} & \textcolor[rgb]{ 0,  .439,  .753}{\textbf{-5.41 }} & -0.24  & 6.03  & -0.68  & 12.94  \\
    \textbf{Sarah} & 0.18  & -3.09  & -2.86  & 59.13  & -2.67  & 55.67  &    & \textcolor[rgb]{ 0,  .439,  .753}{\textbf{2.43 }} & \textcolor[rgb]{ 0,  .439,  .753}{\textbf{-36.06 }} & 1.88  & -44.42  & 1.65  & -36.64  \\
    \rowcolor[rgb]{ .816,  .808,  .808} \textbf{Longdress\_vox10\_1300} & 2.22  & -20.68  & 0.18  & -1.26  & 0.22  & -1.56  &    & 2.03  & -19.34  & \textcolor[rgb]{ 0,  .439,  .753}{\textbf{2.71 }} & \textcolor[rgb]{ 0,  .439,  .753}{\textbf{-33.00 }} & 2.63  & -31.56  \\
    \textbf{Loot\_vox10\_1200} & 1.08  & -14.27  & -6.71  &   314.65$^*$  & -7.33  &  206.02$^*$  &    & \textcolor[rgb]{ 0,  .439,  .753}{\textbf{2.74 }} & \textcolor[rgb]{ 0,  .439,  .753}{\textbf{-37.81 }} & 2.18  & -59.91  & 2.27  & -61.47  \\
    \rowcolor[rgb]{ .816,  .808,  .808} \textbf{Redandblack\_vox10\_1550} & 0.86  & -10.52  & -2.23  & 26.13  & -0.32  & 3.96  &    & 2.48  & -32.77  & 2.23  & -44.28  & \textcolor[rgb]{ 0,  .439,  .753}{\textbf{2.63 }} & \textcolor[rgb]{ 0,  .439,  .753}{\textbf{-32.54 }} \\
    \textbf{Soldier\_vox10\_0690} & 2.22  & -23.82  & -5.96  &   178.44$^*$  & -6.41  &   175.79$^*$  &    & \textcolor[rgb]{ 0,  .439,  .753}{\textbf{3.01 }} & \textcolor[rgb]{ 0,  .439,  .753}{\textbf{-33.13 }} & 2.50  & -57.49  & 2.50  & -57.73  \\
    \rowcolor[rgb]{ .816,  .808,  .808} \textbf{Queen\_0200} & -0.21  & 3.77  & -5.32  &   144.89$^*$  & -4.70  &   127.82$^*$  &    & \textcolor[rgb]{ 0,  .439,  .753}{\textbf{1.15 }} & \textcolor[rgb]{ 0,  .439,  .753}{\textbf{-20.10 }} & 1.02  & -27.84  & 1.12  & -29.66  \\
    \textbf{Arco\_Valentino\_Dense\_vox12} & \textcolor[rgb]{ 0,  .439,  .753}{\textbf{-0.07 }} & \textcolor[rgb]{ 0,  .439,  .753}{\textbf{0.55 }} & -0.37  & 6.23  & -0.35  & 6.16  &    & -3.93  & 37.34  & -2.96  & 38.07  & -3.06  & 37.02  \\
    \rowcolor[rgb]{ .816,  .808,  .808} \textbf{Facade\_00009\_vox12} & \textcolor[rgb]{ 0,  .439,  .753}{\textbf{2.09 }} & \textcolor[rgb]{ 0,  .439,  .753}{\textbf{-18.89 }} & -2.18  & 25.01  & -1.05  & 11.46  &    & 0.22  & -2.04  & 1.61  & -28.39  & 1.01  & -15.98  \\
    \textbf{Egyptian\_mask\_vox12} & \textcolor[rgb]{ 0,  .439,  .753}{\textbf{0.95 }} & \textcolor[rgb]{ 0,  .439,  .753}{\textbf{-13.17 }} & -3.01  & 49.81  & -3.22  & 59.30  &    & 0.24  & -3.71  & 0.61  & -16.28  & 0.70  & -19.71  \\
    \rowcolor[rgb]{ .816,  .808,  .808} \textbf{Shiva\_00035\_vox12} & \textcolor[rgb]{ 0,  .439,  .753}{\textbf{1.17 }} & \textcolor[rgb]{ 0,  .439,  .753}{\textbf{-9.73 }} & -0.44  & 2.70  & -0.91  & 6.97  &    & -1.57  & 14.91  & -0.49  & 6.45  & -0.30  & 4.76  \\
    \textbf{Frog\_00067\_vox12} & \textcolor[rgb]{ 0,  .439,  .753}{\textbf{0.63 }} & \textcolor[rgb]{ 0,  .439,  .753}{\textbf{-7.06 }} & -4.21  & 84.68  & -6.17  &   340.76$^*$  &    & -0.99  & 15.65  & 0.55  & -21.86  & 0.26  & -17.79  \\
    \rowcolor[rgb]{ .816,  .808,  .808} \textbf{Staue\_Klimt\_vox12} & \textcolor[rgb]{ 0,  .439,  .753}{\textbf{1.72 }} & \textcolor[rgb]{ 0,  .439,  .753}{\textbf{-15.11 }} & 0.36  & -2.50  & -0.22  & 3.05  &    & -0.39  & 3.96  & 0.21  & -2.17  & 0.41  & -4.89  \\
    \midrule
    \textbf{Average} & \textbf{0.84 } & \textbf{-8.58 } & \textbf{-2.93 } & \textbf{82.53 } & \textbf{-2.92 } & \textbf{80.69 } &    & \textbf{0.77 } & \textbf{-12.22 } & \textbf{0.85 } & \textbf{-22.56 } & \textbf{0.80 } & \textbf{-19.69 } \\
    \bottomrule
    \bottomrule
    \end{tabular}%
  \label{tab:addlabel}%

       Note that BD-BR$^*$ results in some unreliable results, and BD-PSNR may better reflect the gaps between rate distortion data.
\end{table*}

\begin{table*}[htbp]
  \centering
  \renewcommand\tabcolsep{0.6pt} 
  \renewcommand\arraystretch{0.9}
  \caption{BD-PSNR Comparisons of Case 3 (Triangle Surface Approximation-based Lossy Geometry Compression), the Results of the RAHT Encoder Are Used as the Benchmark.}
    \begin{tabular}{cccccccccccc}
    \toprule
    \toprule
       & \multicolumn{3}{c}{\textbf{LOD encoder w/o LT}} &    & \multicolumn{3}{c}{\textbf{LOD encoder with LT}} &    & \multicolumn{3}{c}{\textbf{TMC2}} \\
\cmidrule{2-4}\cmidrule{6-8}\cmidrule{10-12}    \textbf{    Datasets} & \textbf{Y-BD-} & \textbf{U-BD-} & \textbf{V-BD-} &    & \textbf{Y-BD-} & \textbf{U-BD-} & \textbf{V-BD-} &    & \textbf{Y-BD-} & \textbf{U-BD-} & \textbf{V-BD-} \\
       & \textbf{PSNR (dB)} & \textbf{PSNR (dB)} & \textbf{PSNR (dB)} &    & \textbf{PSNR (dB)} & \textbf{PSNR (dB)} & \textbf{PSNR (dB)} &    & \textbf{PSNR (dB)} & \textbf{PSNR (dB)} & \textbf{PSNR (dB)} \\
\cmidrule{1-4}\cmidrule{6-8}\cmidrule{10-12}    \textbf{Andrew} & 0.08  & -1.50  & -1.96  &    & 0.07  & \textcolor[rgb]{ 0,  .439,  .753}{\textbf{0.29 }} & 0.24  &    & 0.12  & -1.89  & -1.96  \\
    \rowcolor[rgb]{ .816,  .808,  .808} \textbf{Ricardo} & 0.11  & -2.46  & -3.10  &    & 0.73  & \textcolor[rgb]{ 0,  .439,  .753}{\textbf{0.88 }} & 0.76  &    & -0.23  & -1.34  & -1.88  \\
    \textbf{David} & -0.50  & -11.89  & -7.90  &    & 0.68  & \textcolor[rgb]{ 0,  .439,  .753}{\textbf{0.87 }} & 0.84  &    & 0.72  & -1.05  & -1.11  \\
    \rowcolor[rgb]{ .816,  .808,  .808} \textbf{Phil} & 0.09  & -2.77  & -1.68  &    & 0.14  & 0.29  & 0.09  &    & \textcolor[rgb]{ 0,  .439,  .753}{\textbf{0.35 }} & -1.92  & -2.19  \\
    \textbf{Sarah} & 0.21  & -1.79  & -2.01  &    & 0.67  & \textcolor[rgb]{ 0,  .439,  .753}{\textbf{1.39 }} & 1.03  &    & -0.14  & -1.52  & -2.48  \\
    \rowcolor[rgb]{ .816,  .808,  .808} \textbf{Longdress\_vox10\_1300} & 0.74  & -0.68  & -0.56  &    & 0.85  & 2.20  & 1.92  &    & \textcolor[rgb]{ 0,  .439,  .753}{\textbf{3.12 }} & 1.25  & 0.82  \\
    \textbf{Loot\_vox10\_1200} & 0.65  & -4.92  & -7.30  &    & 1.15  & 1.91  & 1.86  &    & \textcolor[rgb]{ 0,  .439,  .753}{\textbf{2.80 }} & 0.57  & 0.49  \\
    \rowcolor[rgb]{ .816,  .808,  .808} \textbf{Redandblack\_vox10\_1550} & 0.37  & -1.60  & -0.19  &    & 1.12  & 1.66  & 0.99  &    & \textcolor[rgb]{ 0,  .439,  .753}{\textbf{2.81 }} & 0.28  & 0.02  \\
    \textbf{Soldier\_vox10\_0690} & 0.84  & -4.98  & -4.18  &    & 1.00  & 2.12  & 2.12  &    & \textcolor[rgb]{ 0,  .439,  .753}{\textbf{2.16 }} & 0.05  & 0.03  \\
    \rowcolor[rgb]{ .816,  .808,  .808} \textbf{Queen\_0200} & 0.25  & -2.49  & -1.95  &    & 0.18  & 0.31  & 0.47  &    & \textcolor[rgb]{ 0,  .439,  .753}{\textbf{1.67 }} & -5.61  & -7.58  \\
    \textbf{Arco\_Valentino\_Dense\_vox12} & 0.02  & -0.69  & -0.40  &    & -0.05  & -0.56  & -0.54  &    & \textcolor[rgb]{ 0,  .439,  .753}{\textbf{2.78 }} & -0.99  & -1.05  \\
    \rowcolor[rgb]{ .816,  .808,  .808} \textbf{Facade\_00009\_vox12} & 0.31  & -1.82  & -0.85  &    & 0.10  & 1.28  & 0.92  &    & \textcolor[rgb]{ 0,  .439,  .753}{\textbf{2.31 }} & -0.01  & -0.67  \\
    \textbf{Egyptian\_mask\_vox12} & \textcolor[rgb]{ 0,  .439,  .753}{\textbf{0.11 }} & -1.06  & -1.76  &    & -0.23  & -0.56  & -0.52  &    & -8.48  & -8.80  & -14.54  \\
    \rowcolor[rgb]{ .816,  .808,  .808} \textbf{Shiva\_00035\_vox12} & 0.06  & 0.00  & -0.10  &    & -0.06  & -0.27  & -0.21  &    & \textcolor[rgb]{ 0,  .439,  .753}{\textbf{2.72 }} & -1.68  & -1.74  \\
    \textbf{Frog\_00067\_vox12} & 0.22  & -3.66  & -4.42  &    & 0.18  & 1.15  & \textcolor[rgb]{ 0,  .439,  .753}{\textbf{0.82 }} &    & -0.22  & -0.47  & -0.47  \\
    \rowcolor[rgb]{ .816,  .808,  .808} \textbf{Staue\_Klimt\_vox12} & 0.08  & 0.06  & -0.13  &    & -0.08  & -0.34  & -0.38  &    & \textcolor[rgb]{ 0,  .439,  .753}{\textbf{3.17 }} & -0.29  & -0.68  \\
    \midrule
    \textbf{Average} & \textbf{0.23 } & \textbf{-2.64 } & \textbf{-2.40 } &    & \textbf{0.40 } & \textbf{0.79 } & \textbf{0.65 } &    & \textbf{0.98 } & \textbf{-1.46 } & \textbf{-2.19 } \\
    \bottomrule
    \bottomrule
    \end{tabular}%
  \label{tab:addlabel}%
\end{table*}%

\begin{table*}[htbp]
  \centering
  \renewcommand\tabcolsep{0.6pt} 
  \renewcommand\arraystretch{0.9}
  \caption{BD-PSNR Comparisons of Case 3 (Direct Geometry Quantization-based Lossy Geometry Compression), the Results of the RAHT Encoder Are Used as the Benchmark.}
    \begin{tabular}{cccccccccccc}
    \toprule
    \toprule
       & \multicolumn{3}{c}{\textbf{LOD encoder w/o LT}} &    & \multicolumn{3}{c}{\textbf{LOD encoder with LT}} &    & \multicolumn{3}{c}{\textbf{TMC2}} \\
\cmidrule{2-4}\cmidrule{6-8}\cmidrule{10-12}    \textbf{    Datasets} & \textbf{Y-BD-} & \textbf{U-BD-} & \textbf{V-BD-} &    & \textbf{Y-BD-} & \textbf{U-BD-} & \textbf{V-BD-} &    & \textbf{Y-BD-} & \textbf{U-BD-} & \textbf{V-BD-} \\
       & \textbf{PSNR (dB)} & \textbf{PSNR (dB)} & \textbf{PSNR (dB)} &    & \textbf{PSNR (dB)} & \textbf{PSNR (dB)} & \textbf{PSNR (dB)} &    & \textbf{PSNR (dB)} & \textbf{PSNR (dB)} & \textbf{PSNR (dB)} \\
\cmidrule{1-4}\cmidrule{6-8}\cmidrule{10-12}    \textbf{Andrew} & 0.19  & -1.20  & -1.24  &    & 0.21  & \textcolor[rgb]{ 0,  .439,  .753}{\textbf{0.64 }} & 0.58  &    & -1.10  & -1.70  & -1.84  \\
    \rowcolor[rgb]{ .816,  .808,  .808} \textbf{Ricardo} & -0.05  & -5.51  & -2.63  &    & \textcolor[rgb]{ 0,  .439,  .753}{\textbf{2.09 }} & 1.77  & 1.89  &    & 0.96  & 0.66  & 0.18  \\
    \textbf{David} & 0.45  & -3.11  & -2.88  &    & 1.13  & \textcolor[rgb]{ 0,  .439,  .753}{\textbf{1.23 }} & 1.18  &    & 0.44  & -0.45  & -0.79  \\
    \rowcolor[rgb]{ .816,  .808,  .808} \textbf{Phil} & 0.28  & -1.65  & -0.84  &    & 0.37  & \textcolor[rgb]{ 0,  .439,  .753}{\textbf{0.88 }} & 0.66  &    & -0.59  & -1.38  & -1.88  \\
    \textbf{Sarah} & 0.52  & -1.31  & -1.15  &    & 0.76  & \textcolor[rgb]{ 0,  .439,  .753}{\textbf{1.47 }} & 1.14  &    & -1.28  & -0.84  & -2.47  \\
    \rowcolor[rgb]{ .816,  .808,  .808} \textbf{Longdress\_vox10\_1300} & 0.83  & -0.17  & -0.13  &    & 0.83  & 2.20  & 1.86  &    & \textcolor[rgb]{ 0,  .439,  .753}{\textbf{2.40 }} & 0.80  & 0.35  \\
    \textbf{Loot\_vox10\_1200} & 0.75  & -6.80  & -7.82  &    & 2.12  & 2.14  & 2.26  &    & \textcolor[rgb]{ 0,  .439,  .753}{\textbf{3.49 }} & 0.92  & 0.93  \\
    \rowcolor[rgb]{ .816,  .808,  .808} \textbf{Redandblack\_vox10\_1550} & 0.82  & -0.39  & 0.18  &    & 1.13  & 1.76  & 0.87  &    & \textcolor[rgb]{ 0,  .439,  .753}{\textbf{1.93 }} & 0.14  & -1.07  \\
    \textbf{Soldier\_vox10\_0690} & 1.42  & -5.09  & -5.54  &    & 1.50  & 2.42  & 2.35  &    & \textcolor[rgb]{ 0,  .439,  .753}{\textbf{1.81 }} & 0.41  & 0.35  \\
    \rowcolor[rgb]{ .816,  .808,  .808} \textbf{Queen\_0200} & 0.16  & -4.01  & -3.40  &    & 0.64  & 0.90  & 0.87  &    & -0.38  & -6.10  & -8.26  \\
    \textbf{Arco\_Valentino\_Dense\_vox12} & \textcolor[rgb]{ 0,  .439,  .753}{\textbf{-0.20 }} & -1.52  & -1.30  &    & -2.94  & -2.06  & -2.11  &    & -11.69  & -3.59  & -3.59  \\
    \rowcolor[rgb]{ .816,  .808,  .808} \textbf{Facade\_00009\_vox12} & \textcolor[rgb]{ 0,  .439,  .753}{\textbf{1.89 }} & -2.96  & -1.67  &    & -0.50  & 1.41  & 0.65  &    & -3.49  & -0.32  & -0.94  \\
    \textbf{Egyptian\_mask\_vox12} & \textcolor[rgb]{ 0,  .439,  .753}{\textbf{0.79 }} & -3.67  & -3.84  &    & 0.28  & 0.64  & 0.73  &    & -17.80  & -11.34  & -16.53  \\
    \rowcolor[rgb]{ .816,  .808,  .808} \textbf{Shiva\_00035\_vox12} & \textcolor[rgb]{ 0,  .439,  .753}{\textbf{1.10 }} & -0.03  & -0.44  &    & -1.64  & -0.58  & -0.37  &    & -11.46  & -5.33  & -4.41  \\
    \textbf{Frog\_00067\_vox12} & \textcolor[rgb]{ 0,  .439,  .753}{\textbf{0.61 }} & -4.09  & -5.99  &    & -0.63  & 0.61  & 0.29  &    & -1.62  & 0.13  & -0.14  \\
    \rowcolor[rgb]{ .816,  .808,  .808} \textbf{Staue\_Klimt\_vox12} & \textcolor[rgb]{ 0,  .439,  .753}{\textbf{1.66 }} & 0.02  & -0.58  &    & -0.30  & 0.43  & 0.63  &    & -7.80  & -3.92  & -3.31  \\
    \midrule
    \textbf{Average} & \textbf{0.70 } & \textbf{-2.59 } & \textbf{-2.45 } &    & \textbf{0.32 } & \textbf{0.99 } & \textbf{0.84 } &    & \textbf{-2.89 } & \textbf{-1.99 } & \textbf{-2.71 } \\
    \bottomrule
    \bottomrule
    \end{tabular}%
  \label{tab:addlabel}%
\end{table*}%

\subsubsection{Triangle surface approximation-based lossy geometry compression}

In this case, \textbf{Appendix B} and \textbf{Appendix C} show the rate and the PSNRs of the TMC13 and TMC2 in detail. The rate-PSNR curves of the four methods are compared in Fig. 11, in which the $x$-axis defines the total bits of geometry and color, while the $y$-axis indicates the Y-PSNRs of the reconstructed point clouds. The average performance, i.e., BD-PSNR is compared in Table IV. We can see that the rate-PSNR curves of TMC2 is usually higher than those of the other methods. It should also be noted that the rate-PSNR curves of TMC2 for the sparse point clouds are not stable (even weird for \textit{Shiva\_00035\_vox12} and \textit{Staue\_Klimt\_vox12}). This reason maybe the sparsity and noise of the \textit{Shiva\_00035\_vox12} and \textit{Staue\_Klimt\_vox12}. In other words, TMC2 is not robust enough for large scale sparse and noise point clouds, although its average performance is better. In the meantime, we can also observe that the performance of the Y component of the \textbf{LoD encoder with LT} is better than that of the \textbf{LoD encoder w/o LT} and \textbf{RAHT encoder}. Strictly speaking, the triangle surface approximation used in the lossily geometry compression of TMC13 can be thought as a local up-sampling process. The color information will be inevitably distorted before compression. This maybe the reason why the objective performance of TMC13 is much lower than TMC2. However, as shown in Fig. 12 and Fig. 13, the gap between TMC13 and TMC2 is not significant when comparing the subjective quality under the similar rate. Specially, we can observe that \textit{David}'s face rebuilt by TMC13 is better than TMC2, but TMC2 works well on the upper body of \textit{David} in Fig. 13. The right shoulder of \textit{Phil} gets serious damage by the encoding of TMC2. Interestingly, for \textit{Loot\_vox10\_1200}, and \textit{Redandblack\_vox10\_1550}, we can hardly see the difference in subjective quality among TMC13 and TMC2.

Moreover, detailed quantitative comparisons of the U and V components are shown in Table IV. It can be observed that the average performance of the \textbf{LoD encoder with LT} is significantly better than the other methods, i.e., the average U-BD-PSNR and V-BD-PSNR of the \textbf{LoD encoder with LT} are 0.79 and 0.65 dB higher than that of the \textbf{RAHT encoder}, which is similar to the results of case 2. Since the lossy geometry compression in TMC13 deteriorates the color information among points, (especially for sparse point clouds), the \textbf{LoD encoder with LT} cannot work well for the sparse point clouds, as we can see that it exhibits a bad BD-PSNR for sparse point clouds in Table IV, i.e., \textit{Arco\_Valentino\_Dense\_vox12}, \textit{Egyptian\_mask\_vox12}, \textit{Shiva\_00035\_vox12}, \textit{Staue\_Klimt\_vox12}.

\begin{figure}[htbp]
\centering
\includegraphics[width=8.8cm,height=6.9cm]{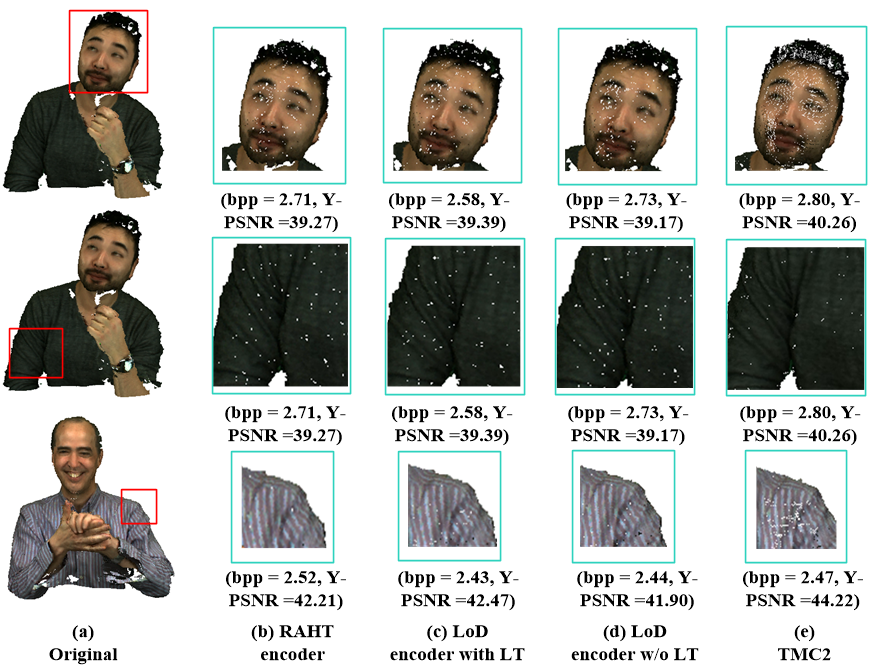}
\caption{Detailed subjective quality comparisons of $David$ and $Phil$.}
\end{figure}

\begin{figure*}[htbp]
\centering
\includegraphics[width=14.7cm,height=10.4cm]{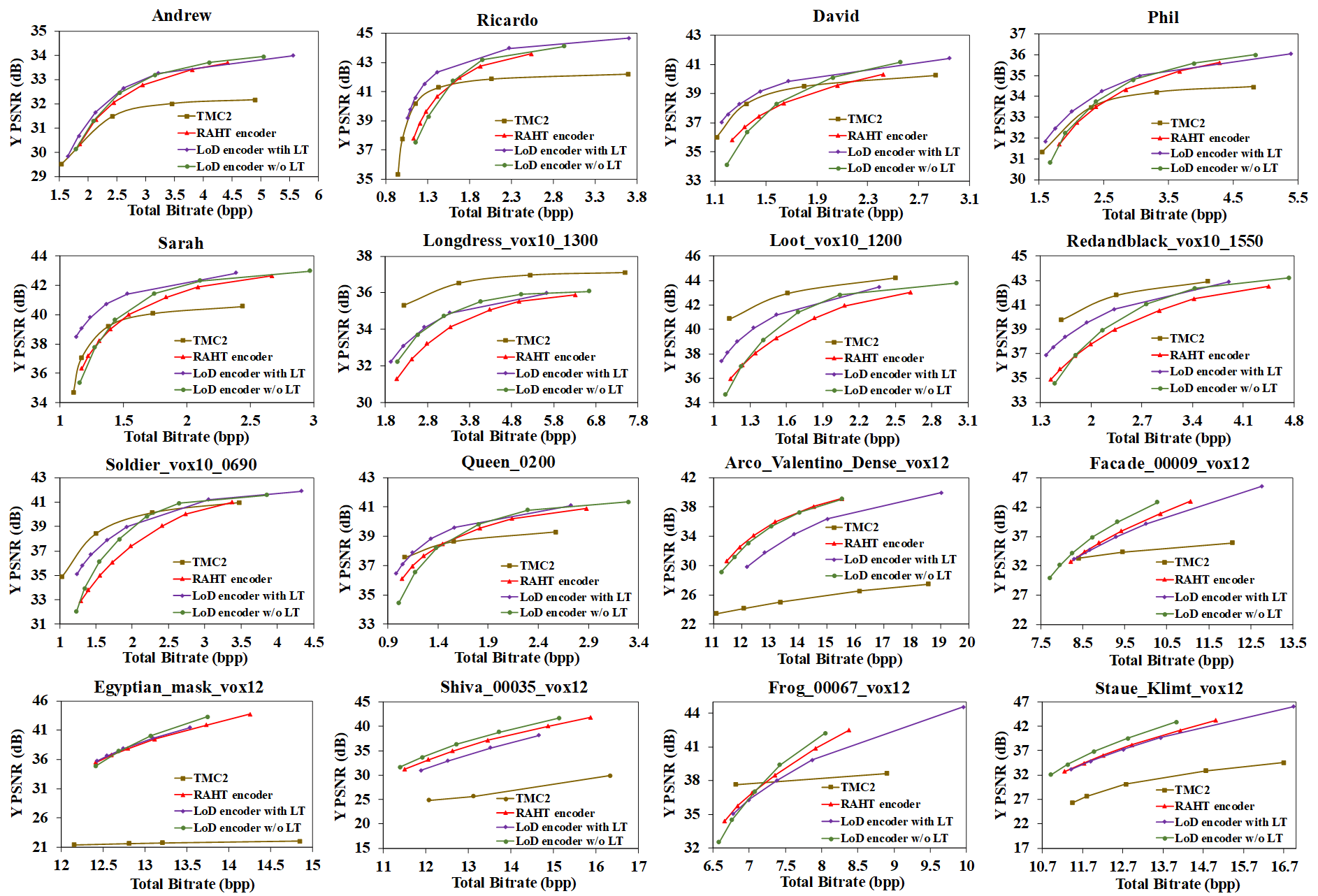}
\caption{Rate-PSNR curves comparison of case 3, in which the direct geometry quantization-based lossy geometry compression is used.}
\end{figure*}

\begin{figure*}[htbp]
\centering
\includegraphics[width=17cm,height=3cm]{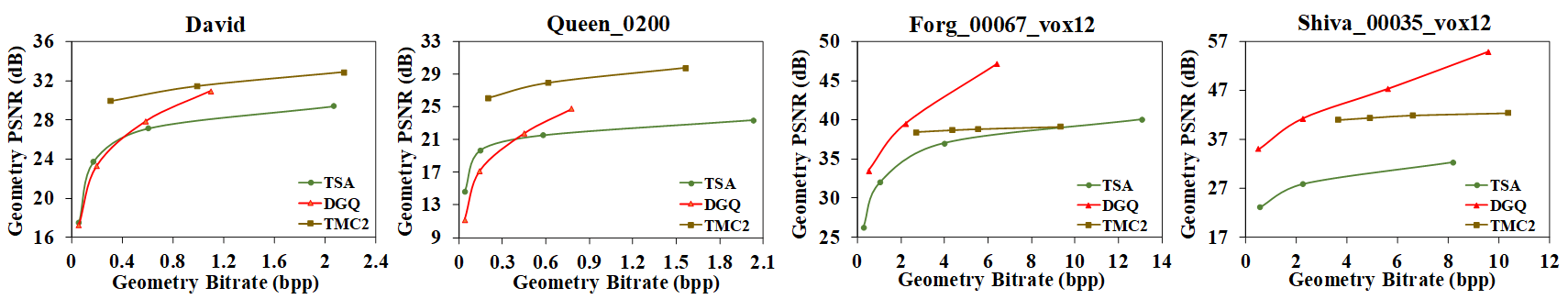}
\caption{Rate-PSNR curves comparison of lossy geometry compression.}
\end{figure*}

\subsubsection{Direct geometry quantization-based lossy geometry compression}

In this case, \textbf{Appendix D} shows the rate and the PSNR data for three methods of TMC13 in detail. For the Y component, the rate-PSNR performance of different methods are compared in Fig. 14 where the $x$-axis refers to the total rate of geometry and color, while, the $y$-axis denotes the Y-PSNRs of reconstructed point cloud. The detailed BD-PSNR is compared in Table V. As we can see in Table V that \textbf{LoD encoder w/o LT} has the best coding efficiency, i.e., an average 0.70 dB BD-PSNR can be achieved compared to the \textbf{RAHT encoder}. However, the coding performance of TMC2 is terrible. The main reason maybe that TMC2 requires to generate a large 2D plane for large scale sparse point clouds (i.e., Fig. 9 (k)-(p)), and thus the correlation of the projected points is broken. Another possible reason is the noise in these point clouds. For the dense point clouds (i.e., Fig. 9 (a)-(j)), the \textbf{LoD encoder with LT} is the better when comparing to the \textbf{RAHT encoder} and the \textbf{LoD encoder w/o LT}, whereas, for the sparse point clouds, the performance of the \textbf{LoD encoder w/o LT} is better.

For the U and V components, compared to the \textbf{RAHT encoder}, an average 0.99, 0.84 dB and the maximum 2.42, 2.35 dB BD-PSNR can be achieved by the \textbf{LoD encoder with LT}, yet the \textbf{LoD encoder w/o LT} and the TMC2 is not better than the \textbf{RAHT encoder}.

\vspace{-2.1ex}

\begin{table}[htbp]
  \centering
  \caption{BD-PSNR Comparisons of Lossy Geometry Compression, the Results of the TSA Method Are Used as the Benchmark.}
  \renewcommand\tabcolsep{3pt} 
  \renewcommand\arraystretch{0.9}
    \begin{tabular}{ccc}
    \toprule
    \toprule
       & \textbf{DGQ} & \textbf{TMC2} \\
    \textbf{Datasets} & \textbf{ BD-PSNR} & \textbf{ BD-PSNR} \\
       & \textbf{(dB)} & \textbf{(dB)} \\
    \midrule
    \textbf{Andrew} & 0.12  & \textcolor[rgb]{ 0,  .439,  .753}{3.14 } \\
    \rowcolor[rgb]{ .816,  .808,  .808} \textbf{Ricardo} & -0.89  & \textcolor[rgb]{ 0,  .439,  .753}{3.50 } \\
    \textbf{David} & -0.13  & \textcolor[rgb]{ 0,  .439,  .753}{3.73 } \\
    \rowcolor[rgb]{ .816,  .808,  .808} \textbf{Phil} & -0.18  & \textcolor[rgb]{ 0,  .439,  .753}{2.86 } \\
    \textbf{Sarah} & 0.13  & \textcolor[rgb]{ 0,  .439,  .753}{3.37 } \\
    \rowcolor[rgb]{ .816,  .808,  .808} \textbf{Longdress\_vox10\_1300} & -1.81  & \textcolor[rgb]{ 0,  .439,  .753}{5.96 } \\
    \textbf{Loot\_vox10\_1200} & -1.61  & \textcolor[rgb]{ 0,  .439,  .753}{6.13 } \\
    \rowcolor[rgb]{ .816,  .808,  .808} \textbf{Redandblack\_vox10\_1550} & -1.48  & \textcolor[rgb]{ 0,  .439,  .753}{4.81 } \\
    \textbf{Soldier\_vox10\_0690} & -1.51  & \textcolor[rgb]{ 0,  .439,  .753}{5.70 } \\
    \rowcolor[rgb]{ .816,  .808,  .808} \textbf{Queen\_0200} & -1.45  & \textcolor[rgb]{ 0,  .439,  .753}{6.29 } \\
    \textbf{Arco\_Valentino\_Dense\_vox12} & \textcolor[rgb]{ 0,  .439,  .753}{17.44 } & 7.65  \\
    \rowcolor[rgb]{ .816,  .808,  .808} \textbf{Facade\_00009\_vox12} & \textcolor[rgb]{ 0,  .439,  .753}{8.51 } & 2.85  \\
    \textbf{Egyptian\_mask\_vox12} & \textcolor[rgb]{ 0,  .439,  .753}{26.34 } & 10.61  \\
    \rowcolor[rgb]{ .816,  .808,  .808} \textbf{Shiva\_00035\_vox12} & \textcolor[rgb]{ 0,  .439,  .753}{14.58 } & 11.19  \\
    \textbf{Frog\_00067\_vox12} & \textcolor[rgb]{ 0,  .439,  .753}{5.05 } & 1.11  \\
    \rowcolor[rgb]{ .816,  .808,  .808} \textbf{Staue\_Klimt\_vox12} & \textcolor[rgb]{ 0,  .439,  .753}{15.40 } & 12.61  \\
    \midrule
    \textbf{Average} & \textbf{4.91 } & \textbf{5.72 } \\
    \bottomrule
    \bottomrule
    \end{tabular}%
  \label{tab:addlabel}%
\end{table}%

\subsubsection{Geometry compression results comparison}

In this case, we compared the rate-distortion performance of lossy geometry compression among three lossy geometry methods (i.e., the triangle surface approximation-based (\textbf{TSA}) method, the direct geometry quantization-based (\textbf{DGQ}) method and TMC2 lossy geometry compression method). The rate-PSNR curves are showed in Fig. 15, in which the $x$-axis denotes the bits of geometry, while, the $y$-axis represents the Geometry-PSNR of the reconstructed point clouds. Corresponding BD-PSNRs are compared in Table VI, when \textbf{TSA} method is used as benchmark. We can observe that TMC2 is better than the other two methods, i.e., an average 5.72 dB BD-PSNR can be achieved by TMC2 compared to the \textbf{TSA} method. The reason is that the \textbf{TSA} method cannot accurately establish the topological surfaces, and thus cannot approximate the original 3D point cloud. Although overall performance of the \textbf{DGQ} method is not the optimal, it is worth pointing out that the \textbf{DGQ} method exhibits an excellent performance among three lossy geometry compression methods for sparse point clouds.

Therefore, we can conclude that TMC2 and the \textbf{LoD encoder with LT} can work well for dense point clouds, but not suitable for sparse point clouds. Although the average performance of the \textbf{RAHT encoder} is not the best, it is robust for all the tested point clouds. Compared with lossy geometry compression of \textbf{DGQ} and \textbf{TSA}, TMC2 exhibits a better coding performance. Note that the performance of \textbf{DGQ} method is much better than \textbf{TSA} method, this is why the overall objective coding performance of case 3.1 is not as good as case 3.2 for TMC13.

\vspace{-2ex}
\subsection{Complexity Analysis}
Regarding to the complexity, we can see that the complexity of TMC2 is much larger than the other methods, as shown in Table II, VII, and VIII. From Table II, the average coding time of TMC13 and TMC2 are 8.28 and 7443.66 seconds respectively, when the geometry and color are losslessly compressed. The high complexity of TMC2 is mainly due to the complexity of the HEVC encoder. From Table VII, we can observe that the \textbf{RAHT encoder} has the smallest coding time compared to the \textbf{LoD encoder with LT} and the \textbf{LoD encoder w/o LT} in case 2. For case 3 in Table VIII, it can also be observed that the encoding time of TMC2 is significantly larger than that of all the other methods.
\begin{table}[htbp]
  \centering
  \caption{The Comparisons of Time Complexity for Case 2.}
  \renewcommand\tabcolsep{3pt} 
  \renewcommand\arraystretch{0.9}
    \begin{tabular}{cccc}
    \toprule
    \toprule
    & {\textbf{RAHT}} & \textbf{Lod encoder } & \textbf{Lod encoder } \\
    \textbf{Datasets} & {\textbf{encoder}}   & \textbf{w/o LT} & \textbf{with LT} \\
\cmidrule{2-4}       & \multicolumn{3}{c}{\textbf{Coding times (s)}} \\
    \midrule
    \textbf{Andrew} & \textcolor[rgb]{ 0,  .439,  .753}{\textbf{2.17 }} & 2.28  & 2.28  \\
    \rowcolor[rgb]{ .816,  .808,  .808} \textbf{Ricardo} & \textcolor[rgb]{ 0,  .439,  .753}{\textbf{1.57 }} & 1.66  & 1.67  \\
    \textbf{David} & \textcolor[rgb]{ 0,  .439,  .753}{\textbf{2.51 }} & 2.71  & 2.81  \\
    \rowcolor[rgb]{ .816,  .808,  .808} \textbf{Phil} & \textcolor[rgb]{ 0,  .439,  .753}{\textbf{2.85 }} & 3.04  & 3.05  \\
    \textbf{Sarah} & \textcolor[rgb]{ 0,  .439,  .753}{\textbf{2.30 }} & 2.52  & 2.44  \\
    \rowcolor[rgb]{ .816,  .808,  .808} \textbf{Longdress\_vox10\_1300} & \textcolor[rgb]{ 0,  .439,  .753}{\textbf{6.70 }} & 7.28  & 7.13  \\
    \textbf{Loot\_vox10\_1200} & \textcolor[rgb]{ 0,  .439,  .753}{\textbf{6.29 }} & 6.65  & 6.69  \\
    \rowcolor[rgb]{ .816,  .808,  .808} \textbf{Redandblack\_vox10\_1550} & \textcolor[rgb]{ 0,  .439,  .753}{\textbf{5.94 }} & 6.25  & 6.27  \\
    \textbf{Soldier\_vox10\_0690} & \textcolor[rgb]{ 0,  .439,  .753}{\textbf{8.72 }} & 9.01  & 9.13  \\
    \rowcolor[rgb]{ .816,  .808,  .808} \textbf{Queen\_vox10\_0200} & \textcolor[rgb]{ 0,  .439,  .753}{\textbf{7.74 }} & 8.25  & 8.28  \\
    \textbf{Arco\_Valentino\_Dense\_vox12} & \textcolor[rgb]{ 0,  .439,  .753}{\textbf{15.51 }} & 16.84  & 16.81  \\
    \rowcolor[rgb]{ .816,  .808,  .808} \textbf{Facade\_00009\_vox12} & \textcolor[rgb]{ 0,  .439,  .753}{\textbf{15.53 }} & 16.62  & 16.75  \\
    \textbf{Egyptian\_mask\_vox12} & \textcolor[rgb]{ 0,  .439,  .753}{\textbf{2.61 }} & 2.89  & 2.94  \\
    \rowcolor[rgb]{ .816,  .808,  .808} \textbf{Shiva\_00035\_vox12} & \textcolor[rgb]{ 0,  .439,  .753}{\textbf{10.03 }} & 11.13  & 11.05  \\
    \textbf{Frog\_00067\_vox12} & \textcolor[rgb]{ 0,  .439,  .753}{\textbf{34.88 }} & 37.19  & 37.46  \\
    \rowcolor[rgb]{ .816,  .808,  .808} \textbf{Staue\_Klimt\_vox12} & \textcolor[rgb]{ 0,  .439,  .753}{\textbf{4.84 }} & 5.28  & 5.34  \\
    \midrule
    \textbf{Average} & \textbf{8.14 } & \textbf{8.72 } & \textbf{8.76 } \\
    \bottomrule
    \bottomrule
    \end{tabular}%
  \label{tab:addlabel}%
\end{table} Compared with the \textbf{TSA} method, the \textbf{DGQ} method has lower complexity in case 3, The main reason is that the triangle surface reconstruction procedure used in the lossy compression of geometry of TMC13 is time consuming.

\vspace{-2ex}
\subsection{Summary}
To sum up, we have the following conclusions: (a) the rate distortion performance of TMC2 is the best on average for dense point clouds, especially for lossy compression, but it is not suitable for large scale sparse and noisy point clouds; (b) among the three encoders in TMC13, i.e., the \textbf{RAHT encoder}, the \textbf{LoD encoder w/o LT}, and the \textbf{LoD encoder with LT}, the rate distortion performance of the \textbf{LoD encoder with LT} is usually the best especially when the \emph{bpp} is low, nevertheless, it is not suitable for compressing sparse point clouds; (c) among all the encoders, the time complexity of the \textbf{RAHT encoder} is the lowest, whereas, that of TMC2 is the highest; (d) the point-to-point quality metric is not efficient to evaluate the subjective quality of the reconstructed point clouds.

\begin{table*}[htbp]
  \centering
  \caption{The Comparisons of Time Complexity for Case 3.}
  \renewcommand\tabcolsep{3pt} 
  \renewcommand\arraystretch{0.9}
    \begin{tabular}{ccccrccccc}
    \toprule
    \toprule
       & \multicolumn{3}{c}{\textbf{Triangle surface approximate}} &    & \multicolumn{3}{c}{\textbf{Direct geometry  quantization}} &    & \textbf{TMC2} \\
\cmidrule{2-4}\cmidrule{6-8}\cmidrule{10-10}       & \textbf{RAHT} & \textbf{LoD encoder } & \textbf{LoD encoder } &    & \textbf{RAHT} & \textbf{LoD encoder } & \textbf{LoD encoder } &    & \textbf{Coding} \\
    \textbf{Datasets} & \textbf{encoder} & \textbf{w/o LT} & \textbf{with LT} &    & \textbf{encoder} & \textbf{w/o LT} & \textbf{with LT} &    & \textbf{Time (s)} \\
\cmidrule{2-4}\cmidrule{6-8}       & \multicolumn{3}{c}{\textbf{Coding Times (s)}} &    & \multicolumn{3}{c}{\textbf{Coding Times (s)}} &    &  \\
    \midrule
    \textbf{Andrew} & 2.16  & 2.25  & 2.25  &    & \textcolor[rgb]{ 0,  .439,  .753}{\textbf{1.83 }} & 1.93  & 1.94  &    & 60.03  \\
    \rowcolor[rgb]{ .816,  .808,  .808} \textbf{Ricardo} & 31.03  & 31.87  & 31.79  &    & \textcolor[rgb]{ 0,  .439,  .753}{\textbf{15.41 }} & 16.64  & 16.91  &    & 64.27  \\
    \textbf{David} & 2.50  & 2.63  & 2.64  &    & \textcolor[rgb]{ 0,  .439,  .753}{\textbf{2.13 }} & 2.29  & 2.28  &    & 61.11  \\
    \rowcolor[rgb]{ .816,  .808,  .808} \textbf{Phil} & 3.39  & 3.41  & 3.49  &    & \textcolor[rgb]{ 0,  .439,  .753}{\textbf{2.59 }} & 2.90  & 2.92  &    & 67.51  \\
    \textbf{Sarah} & 46.48  & 48.88  & 48.61  &    & \textcolor[rgb]{ 0,  .439,  .753}{\textbf{15.30 }} & 16.54  & 16.58  &    & 63.82  \\
    \rowcolor[rgb]{ .816,  .808,  .808} \textbf{Longdress\_vox10\_1300} & 120.06  & 125.30  & 126.85  &    & \textcolor[rgb]{ 0,  .439,  .753}{\textbf{34.38 }} & 37.11  & 37.20  &    & 159.19  \\
    \textbf{Loot\_vox10\_1200} & 7.59  & 7.94  & 7.92  &    & \textcolor[rgb]{ 0,  .439,  .753}{\textbf{5.80 }} & 6.12  & 6.15  &    & 162.63  \\
    \rowcolor[rgb]{ .816,  .808,  .808} \textbf{Redandblack\_vox10\_1550} & 6.97  & 7.24  & 7.28  &    & \textcolor[rgb]{ 0,  .439,  .753}{\textbf{5.43 }} & 5.74  & 5.71  &    & 165.00  \\
    \textbf{Soldier\_vox10\_0690} & 2.87  & 3.00  & 3.01  &    & \textcolor[rgb]{ 0,  .439,  .753}{\textbf{2.45 }} & 2.56  & 2.60  &    & 202.92  \\
    \rowcolor[rgb]{ .816,  .808,  .808} \textbf{Queen\_vox10\_0200} & 7.80  & 8.14  & 8.13  &    & \textcolor[rgb]{ 0,  .439,  .753}{\textbf{6.57 }} & 6.98  & 7.02  &    & 189.36  \\
    \textbf{Arco\_Valentino\_Dense\_vox12} & 6.66  & 6.96  & 6.97  &    & \textcolor[rgb]{ 0,  .439,  .753}{\textbf{5.08 }} & 5.47  & 5.44  &    & 2254.24  \\
    \rowcolor[rgb]{ .816,  .808,  .808} \textbf{Facade\_00009\_vox12} & 1.55  & 1.61  & 1.64  &    & \textcolor[rgb]{ 0,  .439,  .753}{\textbf{1.38 }} & 1.40  & 1.42  &    & 1366.22  \\
    \textbf{Egyptian\_mask\_vox12} & 2.28  & 2.37  & 2.42  &    & \textcolor[rgb]{ 0,  .439,  .753}{\textbf{1.95 }} & 2.04  & 2.06  &    & 3255.60  \\
    \rowcolor[rgb]{ .816,  .808,  .808} \textbf{Shiva\_00035\_vox12} & 20.27  & 21.33  & 21.35  &    & \textcolor[rgb]{ 0,  .439,  .753}{\textbf{9.97 }} & 10.98  & 11.04  &    & 1963.06  \\
    \textbf{Frog\_00067\_vox12} & 9.49  & 9.91  & 10.09  &    & \textcolor[rgb]{ 0,  .439,  .753}{\textbf{7.33 }} & 7.78  & 7.80  &    & 1593.47  \\
    \rowcolor[rgb]{ .816,  .808,  .808} \textbf{Staue\_Klimt\_vox12} & 8.48  & 8.77  & 8.78  &    & \textcolor[rgb]{ 0,  .439,  .753}{\textbf{4.85 }} & 5.29  & 5.37  &    & 835.25  \\
    \midrule
    \textbf{Average} & \textbf{17.47 } & \textbf{18.22 } & \textbf{18.33 } &    & \textbf{7.65 } & \textbf{8.23 } & \textbf{8.28 } &    & \textbf{778.98 } \\
    \bottomrule
    \bottomrule
    \end{tabular}%
  \label{tab:addlabel}%
\end{table*}%

\section{Conclusion And Future Works}
In this paper, we reviewed some basic technologies for point cloud compression as well as the TMC13 (G-PCC) and TMC2 (V-PCC) encoder proposed by MPEG PCC group in detail. At the same time, the performances of the two encoders were compared comprehensively in terms of objective quality (i.e., rate distortion performance), subjective quality and complexity. By observing the comparison results, we concluded that the \textbf{RAHT encoder} is robust for all types of point clouds, while the \textbf{LoD encoder w/o LT} is more efficient in high bit rates, especially for large scale sparse point clouds. Compared to the \textbf{RAHT encoder} and the \textbf{LoD encoder w/o LT}, the \textbf{LoD encoder with LT} has a better performance for dense point clouds, especially in lower bit rates. The TMC2 encoder can achieve the best rate distortion performance on average among all compression settings (especially for lossy geometry and lossy color compression) for dense point clouds. However, it is not suitable for large scale sparse point clouds. The reason maybe that TMC2 requires to generate a large 2D plane for large scale sparse point clouds, and thus, the correlation of the projected points is broken. Another possible reason is the noise in these point clouds. Meanwhile, its time complexity should be optimized further for real-time applications. By introducing and comparing the key technologies of the 3D point cloud encoders proposed by MPEG, we want more researchers to understand the necessity and the current progress of 3D point cloud compression, and encourage them to participate in the research of point cloud compression.

\vspace{-0.3ex}

Although great achievements of point cloud compression have been made in recent years, there are still a large space to further improve the compression efficiency for point clouds. For example, the attribute prediction tools of TMC13 is very limited, and the time complexity of TMC2 is very high. Besides, effective motion prediction methods, efficient entropy encoding methods, rate distortion optimization methods, optimal bit allocation, and rate control technologies for point cloud compression should also be investigated in the near future.


\vspace{-7ex}
%


\begin{IEEEbiography}
[{\includegraphics[width=1in,height=1.15in,clip,keepaspectratio]{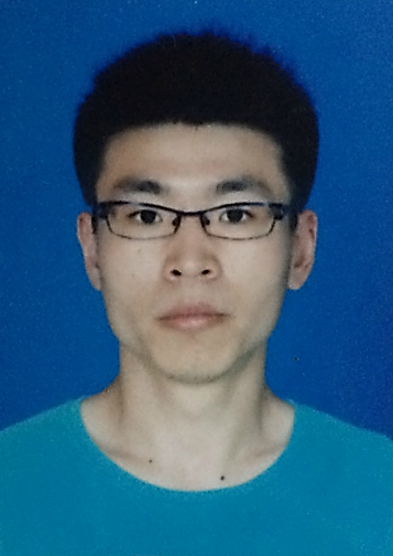}}]{Hao
Liu} received B.S. degree form Department of telecommunication engineering, Shandong Agricultural University, Shandong, China, in 2017. He is currently pursuing the Ph.D. degree  with the Information Science and Engineering from Shandong University. His research interest is point cloud compression and processing.
\end{IEEEbiography}
\vspace{-15ex}

\begin{IEEEbiography}
[{\includegraphics[width=1in,height=1.2in,clip,keepaspectratio]{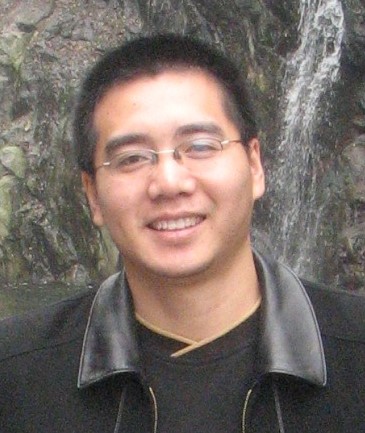}}]{Hui
Yuan} Hui Yuan (S'08--M'12--SM'17) received the B.E. and Ph.D. degree in telecommunication engineering from Xidian University, Xi'an, China, in 2006 and 2011, respectively. From 2011.04 to now, he works as Lecturer, Associate Professor, and Full Professor, Shandong University (SDU), Jinan, China. From 2013-.01-2014.12, he also worked as a post-doctor fellow (Hong Kong Scholar) with the department of computer science, City University of Hong Kong (CityU). His current research interests include video/image/immersive media processing, compression, adaptive streaming, computer vision, etc.
\end{IEEEbiography}

\vspace{-15ex}

\begin{IEEEbiography}
[{\includegraphics[width=1in,height=1.15in,clip,keepaspectratio]{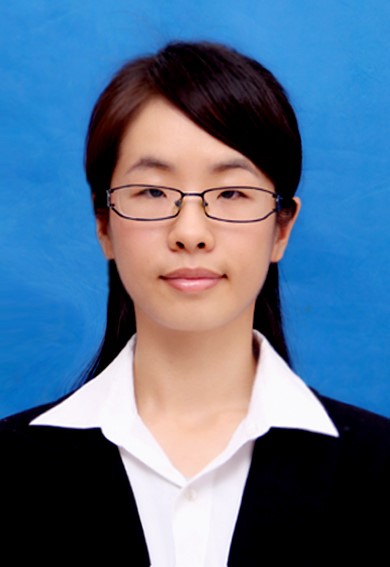}}]{Qi
Liu} received the B.S. degree from Shandong Technology and Business
University, Shandong, China, in 2011 and the M.S. degrees from the
School of telecommunication engineering, Xidian University, Xi'an,
China, in 2014. Currently, she is pursuing the Ph.D. degree with the
Information Science and Engineering from Shandong University,
Shandong, China. Her research interests include point clouds coding
and processing.
\end{IEEEbiography}

\begin{IEEEbiography}
[{\includegraphics[width=1in,height=1.2in,clip,keepaspectratio]{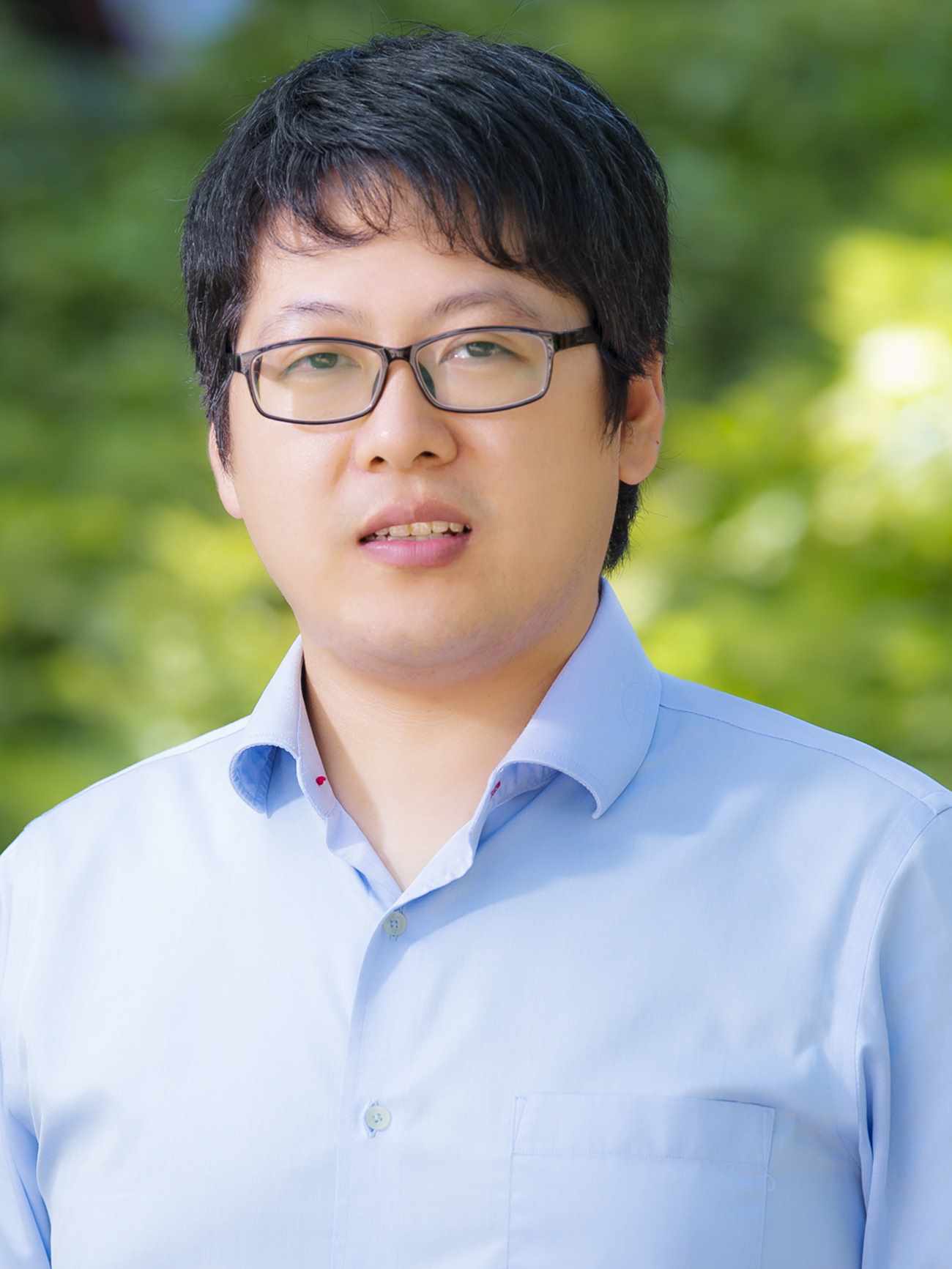}}]{Junhui Hou}(S'13--M'16)
received the B.Eng. degree in information engineering (Talented Students Program)
from the South China University of Technology, Guangzhou, China, in 2009, the M.Eng. degree in signal
and information processing from Northwestern Polytechnical University, Xi'an, China, in 2012, and the
Ph.D. degree in electrical and electronic engineering from the School of Electrical and Electronic
Engineering, Nanyang Technological University, Singapore, in 2016. He has been an Assistant Professor
with the Department of Computer Science, City University of Hong Kong, since 2017. His research
interests fall into the general areas of visual signal processing, such as image/video/3D geometry
data representation, processing and analysis, semi-supervised/unsupervised data modeling for
clustering/classification, and data compression and adaptive transmission.

Dr. Hou was the recipient of several prestigious awards, including the Chinese Government Award
for Outstanding Students Study Abroad from China Scholarship Council in 2015, and the Early Career
Award from the Hong Kong Research Grants Council in 2018. He is serving/served as an Associate
Editor for The Visual Computer, an Area Editor for Signal Processing: Image Communication, the
Guest Editor for the IEEE Journal of Selected Topics in Applied Earth Observations and Remote
Sensing, the Journal of Visual Communication and Image Representation, and Signal
Processing: Image Communication, and an Area Chair of ACM International Conference on Multimedia
(ACM MM) 2019 and IEEE International Conference on Multimedia {\rm{\&}} Expo (IEEE ICME) 2020.
\end{IEEEbiography}
\vspace{-30ex}

\begin{IEEEbiography}
[{\includegraphics[width=1in,height=1.2in,clip,keepaspectratio]{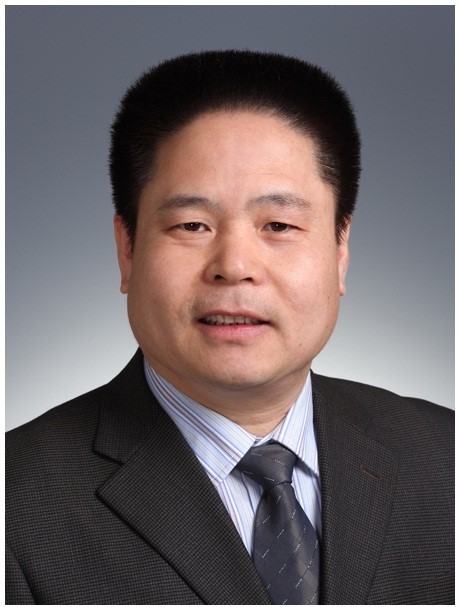}}]{Ju
Liu} (M'02--SM'09) received the B.S. and M.S. degrees in electronics
engineering from Shandong University (SDU), Jinan, China, in 1986
and 1989, respectively, and the Ph.D. degree in signal processing
from the Southeast University (SEU), Nanjing, China, in 2000. Since
July 1989, he has been with the School of Information Science and
Engineering, Shandong University. From July 2002 to December 2003,
he was a visiting professor with the Department of Signal Theory and
Communication, Polytechnic University of Catalonia and the
Telecommunications Technological Center of Catalonia, Barcelona,
Spain. From November 2005 to January 2006, he was a visiting
researcher with the Department of Communications Engineering,
University of Bremen, Bremen, Germany, and the Department of
Communication Systems, University of Duisburg, Essen, Germany. From
June to December, 2009, he was a senior research fellow with the
Department of Electrical Engineering, University of Washington,
Seattle. He is the author or a coauthor of more than 200 journal
papers and conference proceedings. He is a member of editorial
committee of the Journal of Swarm Intelligence Research, the Journal
of Communications, and the Journal of Circuits and Systems in China.
His research interests include space-time processing in wireless
communication, blind signal separation, and multimedia
communications. He received the Program for New Century Excellent
Talents in University of China Award, three Conference Best Paper
awards, and three national and local academic awards in blind signal
processing.
\end{IEEEbiography}


\begin{thebibliography}{00}
\bibitem{b1} S. Schwarz, M. Preda, V. Baroncini, M. Budagavi, P. Cesar, P. A. Chou, R. A. Cohen, M. Krivokuca, S. Lasserre, Z. Li, J. Llach, K. Mammou, R. Mekuria, O. Nakagami, E. Siahaan, A. Tabatabai, A. M. Tourapis, V. Zakharchenko, ``Emerging mpeg standards for point cloud compression,'' \textit{IEEE Journal on Emerging and Selected Topics in Circuits and Systems}, vol. 9, no. 1, pp. 133--148, 2019.
\bibitem{b2} J. Hou, L. Chau, N. Magnenat-Thalmann, and Y. He, ``Compressing 3-D human motions via keyframe-based geometry videos,'' \textit{IEEE Transactions on Circuits and Systems for Video Technology}, vol. 25, no. 1, pp. 51--62, Jan. 2015.
\bibitem{b3} J. Hou, L. Chau, M. Zhang, N. Magnenat-Thalmann, and Y. He, ``A highly efficient compression framework for time-varying 3-D facial expressions,'' \textit{IEEE Transactions on Circuits and Systems for Video Technology}, vol. 24, no. 9, pp. 1541--1553, Sept. 2014.
\bibitem{b4} C. Loop, C. Zhang, and Z. Zhang, ``Real-time high-resolution sparse voxelization with application to image-based modeling,'' \textit{in Proc. HighPerform. Graph. Conf}, 2013, pp. 73--79.
\bibitem{b5} J. Hou, L. Chau, Y. He, M. Zhang, and N. Magnenat-Thalmann, ``Rate distortion model based bit allocation for 3-D facial compression using geometry video,'' \textit{IEEE Transactions on Circuits and Systems for Video Technology}, vol. 23, no. 9, pp. 1537--1541, Sept. 2013.

\bibitem{b6} J. Hou, L. Chau, N. Magnenat-Thalmann, and Y. He, ``Sparse low-rank matrix approximation for data compression,'' \textit{IEEE Transactions on Circuits and Systems for Video Technology}, vol. 27, no. 5, pp. 1043--1054, May. 2017.

\bibitem{b7} J. Hou, L. Chau, N. Magnenat-Thalmann, and Y. He, ``Human motion capture data tailored transform coding,'' \textit{IEEE Transactions on Visualization and Computer Graphics}, vol. 21, no. 7, pp. 848--859, Jun. 2015.

\bibitem{b8} J. Hou, L. Chau, N. Magnenat-Thalmann, and Y. He, ``Low-latency compression of mocap data using learned spatial decorrelation transform,'' \textit{Computer Aided Geometric Design}, vol. 43, pp. 211--225, Mar. 2016.

\bibitem{b9} M. Kaess, R. C. Arkin, and J. Rossignac, ``Compact encoding of robot generated 3d maps for efficient wireless transmission,'' \textit{in IEEE Intl. Conf. on Advanced Robotics, ICAR}, Coimbra, Portugal, Jun. 2003, pp. 324--331.
\bibitem{b10} K. Kohira and  H. Masuda, ``Point-cloud compression for vehicle-based mobile mapping systems using portable network graphics,'' \textit{ISPRS Annals of Photogrammetry, Remote Sensing and Spatial Information Sciences}, vol. 6, pp. 99--106, Sept. 2017.
\bibitem{b11} P. Caillet, Y. Dupuis, ``Efficient lidar data compression for embedded v2i or v2v data handling,'' \textit{arXiv preprint arXiv:1904.05649v1}, Apr. 2019.
\bibitem{b12} ``8i - Real human holograms for AR, VR and MR,'' http://8i.com/.
\bibitem{b13} M. Isenburg, ``LASzip: lossless compression of lidar data,'' \textit{European LiDAR Mapping Forum}, 2012.
\bibitem{b14} https://laszip.org/.
\bibitem{b15} P. A. Chou, O. Nakagami, and E. S. Jang, ``Point cloud compression test model for category 1 v0,'' \textit{ISO/IEC JTC1/SC29/WG11 MPEG}, \textit{N17223}, Macau, Oct. 2017.
\bibitem{b16} K. Mammou, ``PCC test model category 2 v0,'' \textit{ISO/IEC JTC1/SC29/WG11 MPEG}, \textit{N17248}, Macau, Oct. 2017.
\bibitem{b17} K. Mammou, ``PCC test model category 3 v0,'' \textit{ISO/IEC JTC1/SC29/WG11 MPEG}, \textit{N17249}, Macau, Oct. 2017.
\bibitem{b18} K. Mammou, P. A. Chou, ``PCC test model category 13 v2,'' \textit{ISO/IEC JTC1/SC29/WG11 MPEG}, \textit{N17519}, San Diego, Apr. 2018.
\bibitem{b19} R. Schnabel and R. Klein, ``Octree-based point-cloud compression,'' in \textit{Proc. Symp. Point-Based Graph}, Jul. 2006, pp. 111--121.
\bibitem{b20} J. Elseberg, D. Borrmann, A. N{\"u}chter, ``One billion points in the cloud -- an octree for efficient processing of 3D laser scans,'' \textit{ISPRS Journal of Photogrammetry and Remote Sensing}, vol. 76, pp. 76--88, Feb. 2015.
\bibitem{b21} Y. Huang, J. Peng, C. --C. J. Kuo, and M. Gopi, ``A generic scheme for progressive point cloud coding,'' \textit{IEEE Transactions on Visualization and Computer Graphics}, vol. 14, no. 2, pp. 440--453, Mar. 2008.
\bibitem{b22} T. Ochotta and D. Saupe, ``Compression of point-based 3d models by shape-adaptive wavelet coding of multi-height fields,''  in \textit{Proc. Symposium on Point-Based Graphics}, Zurich, Switzerland, Jun. 2004.
\bibitem{b23} J. K. Ahn, K. Y. Lee, J. Y. Sim, and C. S. Kim, ``Large-scale 3d point cloud compression using adaptive radial distance prediction in hybrid coordinate domains,'' \textit{IEEE Journal of Selected Topics in Signal Processing}, vol. 9, no. 3, pp. 422--434, Apr. 2015.
\bibitem{24} P. de Oliveira Rente, C. Brites, J. M. Ascenso and F. Pereira, ``Graph-based static 3d point clouds geometry coding,'' \textit{IEEE Transactions on Multimedia}, vol. 21, no. 2, pp. 284--299, Feb. 2019.
\bibitem{b25} R. Mekuria, K. Blom, and P. Cesar, ``Design, implementation and evaluation of a point cloud codec for tele-immersive video,'' \textit{IEEE Transactions on Circuits and Systems for Video Technology}, vol. 27, no. 4, pp. 828--842, Apr. 2017.
\bibitem{b26} H. Houshiar, A. N{\"u}chter, ``3D point cloud compression using conventional image compression for efficient data transmission,'' in \textit{XXV International Conference on Information, Communication and Automation Technologies (ICAT)}, Sarajevo, Bosnia Herzegovina, Oct. 2015.
\bibitem{b27} C. Zhang, D. Flor\^encio, and C. Loops, ``Point cloud attribute compression with graph transform,'' in \textit{Proc. IEEE Int. Conference on Image Processing}, Paris, France, Sept. 2014, pp. 2066--2070.
\bibitem{b28} R. A. Cohen, D. Tian and A. Vetro, ``Attribute compression for sparse point clouds using graph transforms,'' in \textit{IEEE Int. Conference on Image Processing}, Phoenix, AZ, USA, Sept. 2016, pp. 1374--1378.
\bibitem{b29} Y. Shao, Z. Zhang, Z. Li, K. Fan, and G. Li, ``Attribute compression of 3d point clouds using laplacian sparsity optimized graph transform,'' in \textit{IEEE Visual Communications and Image Processing}, St. Petersburg, Dec. 2017, pp. 1--4.
\bibitem{b30} Y. Shao, Q. Zhang, G. Li, and Z. Li, ``Hybrid point cloud attribute compression using slice-based layered structure and block-based intra prediction,'' \textit{ACM Multimedia}, 2018 (https://arxiv.org/abs/1804.10783).
\bibitem{b31} R. L. de Queiroz and P. A. Chou, ``Compression of 3d point clouds using a region-adaptive hierarchical transform,'' \textit{IEEE Transactions on Image Processing}, vol. 25, no. 8, pp. 3947--3956, Aug. 2016.
\bibitem{b32} J. Hou, L. P. Chau, Y. He, and P. A. Chou, ``Sparse representation for colors of 3d point cloud via virtual adaptive sampling,'' in \textit{IEEE International Conference on Acoustics, Speech and Signal Processing}, New Orleans, Mar. 2017, pp. 2926--2930.
\bibitem{b33} S. Gu, J. Hou, H. Zeng, H. Yuan, K. Ma, ``3D point cloud attribute compression using geometry-guided sparse representation,'' \textit{IEEE Transactions on Image Processing}, vol. 19, pp. 796--808, Aug. 2019.
\bibitem{b34} A. Anis, P. A. Chou, and A. Ortega, ``Compression of dynamic 3d point clouds using subdivisional meshes and graph wavelet transforms,'' in \textit{IEEE International Conference on Acoustics, Speech and Signal Processing}, Shanghai, China, Mar. 2016, pp. 6360--6364.
\bibitem{b35} D. Thanou, P. A. Chou, P. Frossard, ``Graph-based compression of dynamic 3d point cloud sequences,'' \textit{IEEE Transactions on Image Processing}, vol. 25, no. 4, pp. 1765--1778, Apr. 2016.
\bibitem{b36} R. L. de Queiroz and P. A. Chou, ``Motion-compensated compression of dynamic voxelized point clouds,''  \textit{IEEE Trans. Image Process}, vol. 26, no. 8, pp. 3886--3895, Aug. 2017.
\bibitem{b37} L. Li, Z. Li, V. Zakharchenko, J. Chen, and H. Li, ``Advanced 3d motion prediction for video based dynamic point cloud compression,'' \textit{IEEE Transactions on Image Processing}, 2019, doi: 10.1109/TIP.2019.2931621.
\bibitem{b38}  Q. Liu, H. Yuan, J. Hou, H. Liu, R. Hamzaoui, ``Model-based encoding parameter optimization for 3D point cloud compression,'' in \textit{Asia-Pacific Signal and Information Processing Association Annual Summit and Conference (APSIPA ASC)}, Honolulu, USA, Nov. 2018.
\bibitem{b39} J. Elseberg, S. Magnenat, R. Siegwart, A. N{\"u}chter, ``Comparison of nearest--neighbor--search strategies and implementations for efficient shape registration,'' \textit{Journal of Software Engineering for Robotics}, vol. 3, pp. 2--12, Mar. 2012.
\bibitem{b40} J. Schauer, A. N{\"u}chter, ``Collision detection between point clouds using an efficient k--d tree implementation,'' \textit{Advanced Engineering Informatics}, vol. 29, no. 3, pp. 440--458, Aug. 2015.
\bibitem{b41} K. Mammou, P. A. Chou, D. Flynn, M. Krivoku\`ca, O. Nakagami, T. Sugio, ``G-PCC codec description v2,'' \textit{ISO/IEC JTC1/SC29/WG11 MPEG, N18189}, Marrakech, Jan. 2019.
\bibitem{b42} S. Lasserre, D. Flyn, ``Inference of a mode using point location direct coding in tmc3,'' \textit{ISO/IEC JTC1/SC29/WG11 MPEG}, \textit{m42239}, Gwangju, Jan. 2018.
\bibitem{b43} S. Lasserre, D. Flyn, ``Neighbour-dependent entropy coding of occupancy patterns in tmc3,'' \textit{ISO/IEC JTC1/SC29/WG11 MPEG}, \textit{m42238}, Gwangju, Jan. 2018.
\bibitem{b44} S. Lasserre, D. Flyn, ``Intra mode for geometry coding in tmc3,'' \textit{ISO/IEC JTC1/SC29/WG11 MPEG}, \textit{m43600}, Ljubljana, Jul. 2018.
\bibitem{b45} K. Mammou, J. Kim, V. Valentin, F. Robinet, A. Tourapis, Y. Su, ``Efficient implementation of the lifting scheme in tmc13,'' \textit{ISO/IEC JTC1/SC29/WG11 MPEG}, \textit{m43781}, Ljubljana, Jul. 2018.
\bibitem{b46} V. Zakharchenko, ``V-PCC codec description,'' \textit{ISO/IEC JTC1/SC29/WG11 MPEG}, \textit{N18190}, Marrakech, Jan. 2019.

\bibitem{b47} H. Najaf-Zadeh, M. Budagavi, ``Improved point cloud compression efficiency in tmc2 via color smoothing of point cloud prior to texture video generation,'' \textit{ISO/IEC JTC1/SC29/WG11 MPEG}, \textit{m43721}, Ljubljana, Jul. 2018.
\bibitem{b48} D. Graziosi, ``TMC2 patch flexible orientation,'' \textit{ISO/IEC JTC1/SC29/WG11 MPEG}, \textit{m43680}, Ljubljana, Jul. 2018.
\bibitem{b49} D. Graziosi, ``TMC2 optimal texture packing,''  \textit{ISO/IEC JTC1/SC29/WG11 MPEG}, \textit{m43681}, Ljubljana, Jul. 2018.
\bibitem{b50} N. Dawar, H. Najaf-Zadeh, R. Joshi, M. Budagavi, ``PCC tmc2 interleaving in geometry and texture layers,'' \textit{ISO/IEC JTC1/SC29/WG11 MPEG}, \textit{m43723}, Ljubljana, Jul. 2018.
\bibitem{b51} A. Vosoughi, D. Graziosi, A. Tabatabai, ``Point cloud color processing,'' \textit{ISO/IEC JTC1/SC29/WG11 MPEG}, \textit{m42733}, San Diego, Apr. 2018.
\bibitem{b52} C. Guede, J. Ricard, J. Llach, ``Spatially adaptive geometry and texture interpolation,'' \textit{ISO/IEC JTC1/SC29/WG11 MPEG}, \textit{m43658}, Ljubljana, Jul. 2018.
\bibitem{b53} C. Loop, Q. Cai, S. O. Escolano, and P. A. Chou, ``Microsoft voxelized upper bodies - a voxelized point cloud dataset,'' \textit{ISO/IEC JTC1/SC29 Joint WG11/WG1(MPEG/JPEG) input document m38673/M72012}, 2016.
\bibitem{b54} E. d'Eon, B. Harrison, T. Myers, and P. A. Chou, ``8i voxelized full bodies - a voxelized point cloud dataset,'' \textit{ISO/IEC JTC1/SC29 Joint WG11/WG1 (MPEG/JPEG) input document WG11M40059/WG1M74006}, 2017.
\bibitem{b55} S. Schwarz, D. Flynn, ``Common test conditions for point cloud compression,'' \textit{ISO/IEC JTC1/SC29/WG11 MPEG}, \textit{N18175}, Marrakech, Jan. 2019.
\bibitem{b56} http://mpegfs.int-evry.fr/MPEG/PCC/DataSets/pointCloud/CfP/datasets/.
\bibitem{b57} http://mpegx.int-evry.fr/software/MPEG/PCC/TM/.
\bibitem{b58} D. Tian, H. Ochimizu, C. Feng, R. Cohen, and A. Vetro, ``Evaluation metrics for point cloud compression,'' \textit{ISO/IEC JTC1/SC29/WG11 MPEG}, \textit{m39966}, 2017.
\bibitem{b59} G. Bjontegaard, ``Improvements of the bd-psnr model,'' \textit{VCEG-AI11}, Berlin, Germany, Jul. 2008.
\end{thebibliography}
\end{document}